\documentclass[a4paper,twocolumn,11pt]{quantumarticle}
\pdfoutput=1
\usepackage[utf8]{inputenc}
\usepackage[english]{babel}
\usepackage[T1]{fontenc}
\usepackage{amsmath}
\usepackage{amssymb}
\usepackage{hyperref}

\usepackage{tikz}
\usepackage{lipsum}

\usepackage[numbers,sort&compress]{natbib}

\usepackage{afterpage}
\usepackage{braket}
\usepackage{flushend}
\usepackage{mathdots} 
\usepackage{multirow}
\usepackage{siunitx}  
\usepackage{url}
\usepackage{qcircuit}

\usepackage[hang,small,bf]{caption}
\usepackage[subrefformat=parens]{subcaption}
\usepackage{float}
\makeatletter
\let\newfloat\newfloat@ltx
\makeatother
\usepackage{algorithm}
\usepackage{algorithmic}
\renewcommand{\algorithmicensure}{\textbf{Output:}}

\begin{document}

\title{Quantum conjugate gradient method using the positive-side quantum eigenvalue transformation}

\author{Kiichiro Toyoizumi}
\affiliation{Department of Applied Physics and Physico-Informatics, Keio University, Hiyoshi 3-14-1, Kohoku-ku, Yokohama, 223-8522, Japan}
\orcid{0009-0002-0270-0302}
\email{toyoizumi@ppl.appi.keio.ac.jp}
\author{Kaito Wada}
\orcid{0000-0003-4976-4530}
\affiliation{Department of Applied Physics and Physico-Informatics, Keio University, Hiyoshi 3-14-1, Kohoku-ku, Yokohama, 223-8522, Japan}
\author{Naoki Yamamoto}
\orcid{0000-0002-8497-4608}
\affiliation{Department of Applied Physics and Physico-Informatics, Keio University, Hiyoshi 3-14-1, Kohoku-ku, Yokohama, 223-8522, Japan}
\affiliation{Quantum Computing Center, Keio University, Hiyoshi 3-14-1, Kohoku-ku, Yokohama, 223-8522, Japan}
\author{Kazuo Hoshino}
\orcid{0000-0001-5672-9640}
\affiliation{Department of Applied Physics and Physico-Informatics, Keio University, Hiyoshi 3-14-1, Kohoku-ku, Yokohama, 223-8522, Japan}

\maketitle

\begin{abstract}
  Quantum algorithms are still challenging to solve linear systems of equations on real devices. 
  This challenge arises from the need for deep circuits and numerous ancilla qubits.
	We introduce the quantum conjugate gradient (QCG) method using the quantum eigenvalue transformation (QET).
    The circuit depth of this algorithm depends on the square root of the coefficient matrix's condition number $\kappa$, representing a square root improvement compared to the previous 
    quantum algorithms, while the total query complexity worsens.
	The number of ancilla qubits is constant, similar to
	other QET-based algorithms. 
	Additionally, to implement the QCG method efficiently, we devise a QET-based technique that uses only 
	the positive side of the polynomial (denoted by $P(x)$ for $x\in[0,1]$).
	We conduct numerical experiments by applying our algorithm to the one-dimensional Poisson equation
	and successfully solve it.
	Based on the numerical results, our algorithm significantly improves circuit depth, 
	outperforming another QET-based algorithm by three to four orders of magnitude.
\end{abstract}

\section{Introduction}

\afterpage{
\begin{table*}[t]
  \centering
  \caption{
					Comparison of quantum linear system algorithms.
					These algorithms assume efficient access to a given matrix $A$.
					The HHL algorithm and the QSLA using linear combination of unitaries (LCU) 
					assume oracles to access elements of $A$.
					The HHL algorithm with quantum singular-value estimation (QSVE)
					relies on quantum random access memory~\cite{giovannetti2008architectures}.
					The QSVT-based algorithms assume a block encoding of $A$
					with constant ancilla qubits.
					Only our algorithm assumes a positive definite Hermitian matrix $A$. 
					The maximum circuit depths of the algorithms, excluding the HHL algorithm with
					variable time amplitude amplification (VTAA) and
					eigenstate filtering by QSVT with quantum Zeno effect (QZE),
					are derived without amplitude amplification.
					On the other hand, the total query complexities of the algorithms, excluding 
					the eigenstate filtering by QSVT with QZE and our algorithm, are derived with amplitude amplification.
					The parameter $\kappa$ is the condition number of $A$, $\varepsilon$ is 
					the desired precision, and $\| \cdot\|_{F}$ is the Frobenius norm.
					The parameters $c_1$ and $c_2$ are related to the dependence of 
					the maximum absolute values of polynomials (see the details in Sec.~\ref{subsec:QCG_method}). 
					}
  \scalebox{0.92}{
		\begin{tabular}{cccc} \hline \hline
      Algorithm & Maximum circuit depth & No. of ancilla qubits & Total query complexity \\ \hline 
			HHL~\cite{harrow2009quantum} & $\tilde{O}(\kappa/\varepsilon)$ & $O(\mathrm{polylog}(\kappa/\varepsilon))$ & $\tilde{O}(\kappa^2/\varepsilon)$ \\      
      HHL with VTAA~\cite{ambainis2010variable} & $\tilde{O}(\kappa/\varepsilon)$ & $O(\mathrm{polylog}(\kappa/\varepsilon))$ & $\tilde{O}(\kappa/\varepsilon)$ \\     
      LCU~\cite{childs2017quantum} & $\tilde{O}(\kappa\mathrm{polylog}(1/\varepsilon))$ & $O(\log(\kappa/\varepsilon))$ & $\tilde{O}(\kappa^2\mathrm{polylog}(1/\varepsilon))$ \\     
      HHL with QSVE~\cite{wossnig2018quantum} & $\tilde{O}(\kappa\lVert A\rVert _{F}/\varepsilon)$ & $O(\mathrm{polylog}(\kappa/\varepsilon))$ & $\tilde{O}(\kappa^2\lVert A\rVert _{F}/\varepsilon)$ \\ 
      \begin{tabular}{c}
				Eigenstate filtering \\by QSVT with QZE~\cite{lin2020optimal}
			\end{tabular} & $\tilde{O}(\kappa\log(1/\varepsilon))$ & $O(1)$ & $\tilde{O}(\kappa\log(1/\varepsilon))$   \\     
      Direct QSVT~\cite{martyn2021grand} & $\tilde{O}(\kappa\log(1/\varepsilon))$ & $O(1)$ & $\tilde{O}(\kappa^2\log(1/\varepsilon))$  \\     
      \begin{tabular}{c}
				QCG by QET \\(our work)
			\end{tabular}& $\tilde{O}\left(\sqrt{\kappa}\log\left(\frac{\|\ket{b}\|}{\|A\|\varepsilon}\right)\right)$ & $O(1)$ & $\tilde{O}\left(\kappa^{5+c_1}\left(\frac{\|\ket{b}\|}{\|A\|\varepsilon}\right)^4+\kappa^{1/2+c_2}\right)$ \\ \hline \hline  
    \end{tabular}
    }
  \label{tb:comparison_of_QLSAs}
\end{table*}
}

\subsection{Background} \label{subsec:background}
Linear systems of equations are found in various fields, such as physics, engineering,
and machine learning. The standard method for solving large systems involves using a computer.
However, solving these systems demands a significant amount of computation time because
the time complexity of the standard algorithms scales as $\mathrm{poly}(N)$ for
the system size $N$. A promising alternative approach is
the use of a quantum computer.
The first quantum algorithm for solving linear systems of equations is the
Harrow-Hassidim-Lloyd (HHL) algorithm~\cite{harrow2009quantum}.
Under specific conditions,
the HHL algorithm achieves an exponential speedup over classical algorithms,
such as the conjugate gradient (CG) method~\cite{shewchuk1994introduction,saad2003iterative}.
This discovery has led to various quantum linear system algorithms (QLSAs)~\cite{ambainis2010variable,
childs2017quantum,wossnig2018quantum,shao2018quantum,
gilyen2019quantum,lin2020optimal,martyn2021grand}.

However, the practical implementation of these algorithms on real devices
is anticipated to be difficult for the next few decades.
This challenge arises due to the need for deep circuits and numerous ancilla qubits.
For instance, the HHL algorithm, relying on quantum phase estimation~\cite{nielsen2010quantum}, 
requires deep circuits with many controlled time evolution operators.
The HHL algorithm achieves linear dependence of circuit depth on the condition number $\kappa$
of a given matrix through variable time amplitude amplification~\cite{ambainis2010variable}.
However, implementing this algorithm on real devices remains difficult.
Moreover, because of the need to store the eigenvalues of the matrix 
in ancilla qubits with a certain precision, the number of ancilla qubits
depends on $\kappa$ and a desired precision $\varepsilon$,
resulting in numerous ancilla qubits.
An effective approach to avoid this dependence is
using the quantum singular-value transformation (QSVT)~\cite{gilyen2019quantum}.

The QSVT is a quantum algorithm that calculates a polynomial of a given matrix 
embedded in a larger unitary matrix.
Notably, the QSVT provides a unified approach to major quantum algorithms, such as 
the Grover search~\cite{grover1996fast}, quantum phase estimation, quantum walks~\cite{szegedy2004quantum},
Hamiltonian simulation~\cite{feynman1982simulating,lloyd1996universal}, 
and matrix inversion.
In the QSVT-based QLSAs~\cite{gilyen2019quantum,lin2020optimal,martyn2021grand}, 
the number of ancilla qubits is constant, regardless of both $\kappa$ and $\varepsilon$. 
However, reducing the circuit depth of these algorithms is desirable in terms of their implementation on real devices.
Therefore, our goal is to present a QSVT-based QLSA 
with shallow circuits.

\subsection{Contribution of this paper}
We define some terms and notations to describe the contributions of this paper. Let $A$ and $\ket{b}$ be
a given invertible $N\times N$ matrix and $N$-dimensional vector; then 
$\ket{x}$ is the solution of a linear system of equations $A\ket{x}=\ket{b}$. Note that the vectors $\ket{x}$ and $\ket{b}$ are not normalized.
The condition number of $A$ is defined by $\kappa=\|A\|\|A^{-1}\|$, where
$\|\cdot\|$ represents the $L_2$ norm in this paper. 
Suppose that $A$ can be efficiently accessed through an oracle or a block encoding (defined in Sec.\ref{subsec:QET})
and $\ket{b}/\|\ket{b}\|$ can be efficiently prepared through an oracle. 
Then the QLSA is a quantum algorithm that produces a quantum state $\ket{\tilde{x}}/\|\ket{\tilde{x}}\|$
such that $\|\ket{x}-\ket{\tilde{x}}\|\leq \varepsilon$. The circuit depth is defined by the number of queries to the block encoding of $A$ in the QSVT framework or the oracle to access elements of $A$ in others.
The maximum circuit depth is the highest depth among all the circuits in the algorithm.
Total query complexity is the overall number of queries to obtain the state $\ket{\tilde{x}}/\|\ket{\tilde{x}}\|$, expressed as 
the sum of the products of the circuit depths and the number of circuit runs.
We use the asymptotic notation $O(t\mathrm{polylog}(t))=\tilde{O}(t)$.
The $\varepsilon$ approximation of a function $f$ to $g$ on the domain $I$ is 
denoted by the inequality $\max_{x\in I}|g(x)-f(x)|\leq\varepsilon.$

The CG method~\cite{shewchuk1994introduction,saad2003iterative}
is a well-established classical algorithm for solving a linear system of equations
involving a positive definite Hermitian matrix $A$.
In this method, vectors are represented as the product of
the polynomial $P(A)$ and the initial vector $\ket{b}$. 
Based on this observation, we introduce the quantum conjugate gradient (QCG) method
using the quantum eigenvalue transformation (QET), corresponding to 
the QSVT for a Hermitian matrix. 
This algorithm iteratively constructs new polynomials and produces corresponding states
in each iteration. In contrast, the QSVT-based QLSA~\cite{gilyen2019quantum,martyn2021grand} 
constructs a polynomial directly approximating $A^{-1}$.
Thus, we call this algorithm the QLSA using the direct QSVT.
Table~\ref{tb:comparison_of_QLSAs} compares the QLSAs regarding
the maximum circuit depth, the number of ancilla qubits, and the total query complexity.

Our algorithm achieves a square-root improvement for $\kappa$ in
the maximum circuit depth.
This enhancement stems from the convergence of the CG method. In the CG method,
the number of iterations required to meet the convergence criterion scales
as $m=O\left(\sqrt{\kappa}\log\left(\frac{\kappa\|\ket{b}\|}{\|A\|\varepsilon}\right)\right)$
The maximum degree of the polynomial $P(A)$ is represented as $m+1$.
Since our algorithm estimates inner products using two polynomials,
the maximum circuit depth is $2(m+1)=O\left(\sqrt{\kappa}\log\left(\frac{\kappa\|\ket{b}\|}{\|A\|\varepsilon}\right)\right)$.
Furthermore, the number of ancilla qubits 
remains constant, independent of both $\kappa$ and $\varepsilon$, because the QET framework formulates our algorithm.
Therefore, our algorithm achieves shallow circuits with a constant number of ancilla qubits.

Note that the total query complexity of our algorithm is less favorable compared to that of other QLSAs.
This is due to the necessity of numerous circuit runs 
to estimate inner products using swap tests~\cite{buhrman2001quantum}.
This trade-off implies that our algorithm achieves shallow circuits at the expense of 
increasing the number of circuit runs.
However, the total query complexity is derived as a worst case and can be improved by optimizing the number of circuit runs and the precision for estimating inner products.

We show that the polynomials' maximum absolute values of the negative side ($x\in[-1,0]$) can grow significantly in specific problems where the QCG method is applied.
Addressing this growth requires amplitude amplification that demands an enormous cost in the previous QET framework.
To overcome this challenge, we introduce a novel technique named positive-side QET.
This technique allows us to use only the positive side ($x\in[0,1]$) of a polynomial for 
a positive semidefinite Hermitian matrix $A$. Importantly, this technique requires 
neither amplitude ampliﬁcation nor additional cost, assuming the block encoding of
$A'=2A-I$ exists. 
The QCG method can be implemented efficiently by integrating this technique.
Additionally, we elaborate on the explicit construction of this block encoding 
to ensure the applicability of the positive-side QET.
In the QET framework, certain constraints impose limitations on the range of 
applicable polynomials. Specifically, the absolute value
of polynomials for $x\in[-1,1]$ must not exceed $1$.
Accordingly, polynomials that meet the constraints have been chosen, such as 
the Chebyshev polynomial and approximate polynomials of trigonometric and sign functions.~\cite{gilyen2019quantum,martyn2021grand}.
The positive-side QET can efficiently handle polynomials with an enormous absolute value on the negative side.
Importantly, the application of this technique is not limited to the QCG method; it can be applied more broadly.
Therefore, this technique broadens the scope of applicable polynomials in the QET framework.

We apply the QCG method using the positive-side QET to solve the one-dimensional Poisson equation for testing.
We achieve successful solutions using the Statevector Simulator provided by Qiskit~\cite{Qiskit}. 
Notably, our algorithm's maximum circuit depth
is significantly smaller (three to four orders) than that of the QLSA using the direct QSVT~\cite{martyn2021grand}. 
Thus, we numerically demonstrate that our algorithm exhibits shallow circuits.


\subsection{Related works}
The QLSAs have evolved toward improving their dependence on $\kappa$ and $\varepsilon$.
The total query complexity of the HHL algorithm~\cite{harrow2009quantum} is $\tilde{O}(\kappa^2/\varepsilon)$.
For any $\delta > 0$, a QLSA with cost $O(\kappa^{1-\delta})$ would imply BQP=PSPACE.
Therefore, the lower bound of the QLSA's total query complexity is regarded as $\Omega(\kappa)$.
Note that this fact is consistent with the square root dependence for $\kappa$ of our algorithm's maximum circuit depth because the dependence of the total query complexity is superlinear.
Ambainis developed variable time amplitude amplification (VTAA) to
improve the dependence from $\kappa^2$ to $\kappa$~\cite{ambainis2010variable}; 
however, the $1/\varepsilon$-dependence remains polynomial. 
In~\cite{childs2017quantum}, the authors reduced this dependence
to log-polynomial using the method for implementing linear combinations of unitaries (LCU). 
The LCU approach requires $O(\log d)$ ancilla qubits, where $d$ is 
the degree of approximation. Since this approach utilizes an approximation
of $1/x$, the number of ancilla qubits depends on $\kappa$ and $\varepsilon$. 
In~\cite{wossnig2018quantum}, the authors presented the QLSA for dense matrices 
by combining the HHL with the quantum singular-value estimation (QSVE)~\cite{kerenidis2016quantum}.
In~\cite{gilyen2019quantum}, the authors introduced the QSVT and explained the QLSA
in this framework. The details were provided in~\cite{martyn2021grand}.
The maximum circuit depth of this algorithm without amplitude amplification scales as
$O(\kappa\log(\kappa/\varepsilon))$, which represents a better scaling 
compared to other QLSAs~\cite{ambainis2010variable,childs2017quantum,wossnig2018quantum,
shao2018quantum,lin2020optimal}.
However, in the worst case, the total query complexity with amplitude amplification scales as
$O(\kappa^2\log(\kappa/\varepsilon))$.
In~\cite{lin2020optimal}, the authors proposed eigenstate filtering by the QSVT.
They reduced the dependence from $\kappa^2$ to $\kappa$
in the total query complexity by combining this method and 
quantum adiabatic computing~\cite{albash2018adiabatic} or quantum Zeno effect~\cite{boixo2009eigenpath}.
Specifically, the query complexity of the method with the quantum Zeno effect 
is $O(\kappa\left[\log(\kappa)\log\log(\kappa) + \log(1/\varepsilon)\right])$.
The number of ancilla qubits of these QSVT-based QLSAs
is constant, regardless of both $\kappa$ and $\varepsilon$.


The QCG method was initially discussed by Shao~\cite{shao2018quantum}. 
The author also focused on the representation of vectors in the CG method, expressed
as the product of the polynomial $P(A)$ and the initial vector $\ket{b}$.
A technique based on the quantum phase estimation was developed 
to prepare an approximate linear combination of two quantum states.
This technique is repeatedly applied to construct the polynomial, resulting in 
the number of ancilla qubits dependent on the polynomial degree.
The maximum degree is determined by $\kappa$ and $\varepsilon$;
therefore, the number of ancilla qubits depends on both. 
Moreover, the scaling for $\kappa$ of the maximum circuit depth can be superlinear. 
Consequently, Shao's QCG method does not achieve the shallow circuit with 
constant ancilla qubits.

The CG method is a form of the Krylov subspace method.
Various quantum Krylov subspace methods have been developed~\cite{parrish2019quantum,
motta2020determining,stair2020multireference,seki2021quantum,cortes2022quantum,kirby2023exact}.
These algorithms are designed to address eigenvalue problems, particularly 
finding the smallest eigenvalue rather than linear systems of equations.
In these algorithms, a Hamiltonian eigenvalue problem is transformed into a generalized
eigenvalue problem. The resolution of these problems involves the evaluation of inner products
and expectation values of states in the Krylov subspace.
The states are generated through operations such as
the real-time evolution 
operators~\cite{parrish2019quantum,stair2020multireference,seki2021quantum,cortes2022quantum},
the imaginary-time evolution operators~\cite{motta2020determining},
or the Chebyshev polynomial~\cite{kirby2023exact}.


\section{Preliminary} \label{sec:preliminary}

\subsection{Quantum eigenvalue transformation} \label{subsec:QET}
The quantum eigenvalue transformation~\cite{gilyen2019quantum,martyn2021grand,low2019hamiltonian} is a quantum algorithm
for calculating a polynomial transformation $P(A)$, where $P$ is a polynomial, and 
$A$ is a Hermitian matrix embedded in a larger unitary matrix.
The QET is derived from quantum signal processing~\cite{low2017optimal,low2019hamiltonian}
and consists of three steps:
input, processing, and output. The input involves a block encoding, representing a square matrix as
the upper left block of a unitary matrix. The block encoding is defined as follows.
Let $n_s$ and $n_a$ be the number of the system and ancilla qubits.
Let $A$ be a Hermitian matrix acting on $n_s$ qubits and $U_A$ be a unitary matrix acting on $n_a+n_s$ qubits.
Then for a subnormalization factor $\alpha > 0$ and an error $\varepsilon > 0$,
$U_A$ is called $(\alpha, n_a, \varepsilon)$ block encoding of $A$ if 
\begin{equation}
  \left\|
    A - \alpha(\bra{0}_a\otimes I_s)U_A(\ket{0}_a\otimes I_s)
  \right\| \leq \varepsilon,
\end{equation}
where $\ket{0}_a = \ket{0}^{\otimes n_a}$ and $I_s$ is an identity matrix acting on $n_s$ qubits. 
Note that we necessarily have $\|A\|\leq\alpha+\varepsilon$ because $\|U\|=1$.

The QET processes the input $U_A$ using the following sequence with $d+1$ real parameters $\Phi=(\phi_0,\phi_1,\dots,\phi_d)$, 
known as a phase factor: For even $d$,
\begin{equation} \label{eq:alternating_phase_modulation_sequence_for_even_polynomial}
  U_{\Phi} \equiv e^{i\phi_0\Pi}\prod_{k=1}^{d/2}\left(
      U_{A}^{\dagger} e^{i\phi_{2k-1}\Pi} U_{A} e^{i\phi_{2k}\Pi}
    \right),
\end{equation}
for odd $d$,
\begin{equation} \label{eq:alternating_phase_modulation_sequence_for_odd_polynomial}
  U_{\Phi} \equiv e^{i\phi_0\Pi} U_A e^{i\phi_1\Pi}\prod_{k=1}^{(d-1)/2}\left(
      U_{A}^{\dagger} e^{i\phi_{2k}\Pi}U_A e^{i\phi_{2k+1}\Pi}
      \right),
\end{equation}
where $\Pi = 2\ket{0}_a\bra{0} - I$. The operator $e^{i\phi\Pi}$ changes the input using
the angle of the phase factor.

\begin{figure*}[tbp]
	\begin{minipage}[c]{\linewidth}
		\centering
    \includegraphics[width=\linewidth]{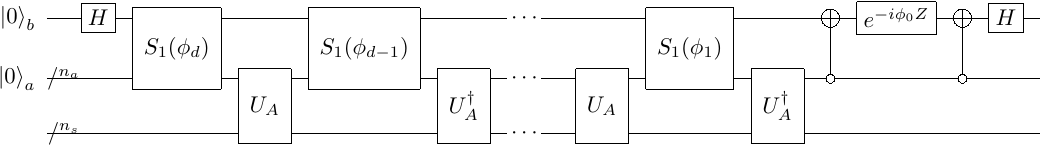}
		\put(-500,75){\Large{(a)}}
		\label{fig:QSVT_circuit_for_even_polynomial}
	\end{minipage} \\ 
	\begin{minipage}[c]{\linewidth}
		\centering
    \includegraphics[width=\linewidth]{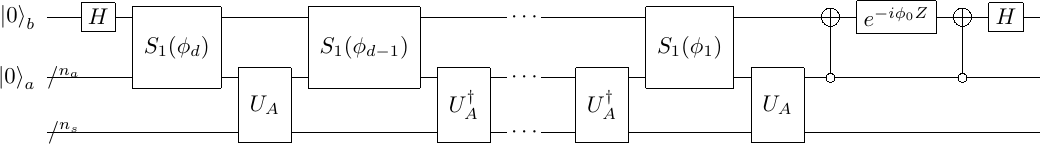}
		\put(-500,75){\Large{(b)}}
		\label{fig:QSVT_circuit_for_odd_polynomial}
	\end{minipage}
	\caption{The QET circuits for a degree-$d$ real polynomial with definite parity:
					(a) For an even polynomial and (b) for an odd polynomial.
					}
	\label{fig:QSVT_circuit_for_polynomial_with_definite_parity}
\end{figure*}

\begin{figure}[tbp]
	\centering
  \includegraphics[width=0.95\linewidth]{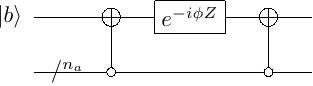}
	\caption{The gate $S_1(\phi)$. For $b=\{0,1\}$, this gate is expressed as 
					$S_1(\phi) = \sum_{b}\ket{b}\bra{b}\otimes e^{(-1)^b i\phi\Pi}$.
					The anti-controlled NOT gate is controlled by $n_a$ qubits.}
	\label{fig:circuit_of_one_phi}
\end{figure}

The main result of the QET is that the output is represented as
$(1,n_a+1,0)$ block encoding of $P(A/\alpha)$ if $U_A$ is $(\alpha, n_a, \varepsilon)$ block encoding of $A$ and a degree-$d$ real polynomial $P\in\mathbb{R}[x]$ satisfies
the conditions
\begin{itemize}
  \item[(i)] $P$ has parity $d \bmod{2}$,
  \item[(ii)] $\forall x\in[-1,1]: |P(x)|\leq 1$.
\end{itemize}
The phase factor $\Phi$ for the even or odd polynomial can be calculated classically in time $O(\mathrm{poly}(d))$
from the polynomial coefficients~\cite{haah2019product,chao2020finding,
dong2021efficient,wang2022energy,ying2022stable,dong2022infinite}.
Figure~\ref{fig:QSVT_circuit_for_polynomial_with_definite_parity}
illustrates the QET circuits for a degree-$d$ real polynomial $P$ with definite parity, representing
$(1, n_a+1, 0)$ block encoding of $P(A/\alpha)$.
If $P$ is a polynomial that $\varepsilon$ approximates a given function $f$ for $x\in[-1,1]$, then
the circuit becomes $(1, n_a+1, \varepsilon)$ block encoding of $f(A/\alpha)$.
The QET circuit features the gate $S_1(\phi)$ shown in Fig.~\ref{fig:circuit_of_one_phi},
which has one angle of the phase factor.
This gate is essentially equal to the processing operator $e^{i\phi\Pi}$ in 
Eqs.~\eqref{eq:alternating_phase_modulation_sequence_for_even_polynomial}
and~\eqref{eq:alternating_phase_modulation_sequence_for_odd_polynomial}.
Since the circuit includes just $d$ queries to $U_A$, the polynomial degree 
is regarded as a metric for the QET circuit's depth.

\subsubsection{Approximate polynomials of sign and rectangular functions} \label{subsubsec:approximate_polynomials_of_sign_and_rectangular_functions}
Let $\delta\in(-1,1), \Delta>0$, and $\varepsilon\in(0, \sqrt{8/e\pi}]$.
According to the results of~\cite{low2017quantum,mitarai2023perturbation},
there exists a polynomial $P_{\delta, \Delta, \varepsilon}^{\mathrm{sgn}}(x)$ that $\varepsilon$ approximates a sign function
\begin{equation}
  \mathrm{sgn}(x-\delta) = \begin{cases}
    -1 & x<\delta \\
    0  & x=\delta \\
    1  & x>\delta,
  \end{cases}
\end{equation}
for $x\in[-1,\delta-\frac{\Delta}{2}]\cup [\delta+\frac{\Delta}{2},1]$ with the degree
\begin{align}
  &d_{\mathrm{sgn}}(\delta, \Delta, \varepsilon) \notag \\
    &= 2\left\lceil
      \frac{16(1+|\delta|)k}{\sqrt{\pi}\varepsilon}\exp\left[
        -\frac{1}{2}W\left(\frac{512}{\pi\varepsilon^{2} e^2}\right)
      \right]
    \right\rceil + 1 \notag \\
    &=O\left(
      \frac{1+|\delta|}{\Delta}\log(1/\varepsilon)
    \right), 
\end{align}
where $W$ is the Lambert $W$ function and
\begin{equation}
	k(\Delta, \varepsilon)=\frac{1}{\Delta}\sqrt{2\ln\left(\frac{8}{\pi\varepsilon^{2}}\right)}.
\end{equation}

\begin{figure*}[t]
	\centering
	\begin{minipage}[c]{\linewidth}
		\centering
    \includegraphics[width=\linewidth]{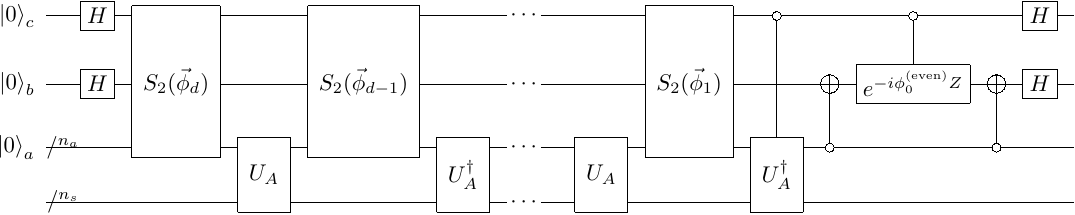}
		\put(-500,105){\Large{(a)}}
		\label{fig:QSVT_circuit_for_general_polynomial_for_even_d}
	\end{minipage} \\ 
	\begin{minipage}[c]{\linewidth}
		\centering
    \includegraphics[width=\linewidth]{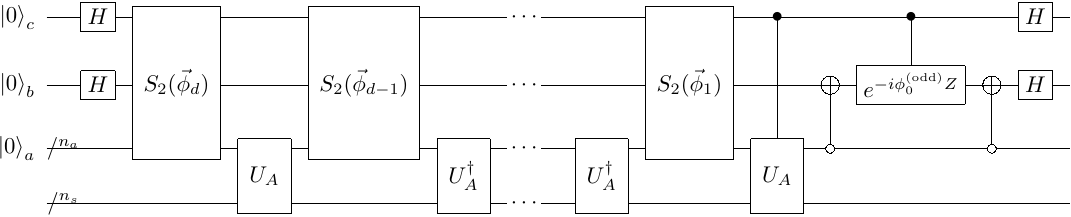}
		\put(-500,105){\Large{(b)}}
		\label{fig:QSVT_circuit_for_general_polynomial_for_odd_d}
	\end{minipage}
	\caption{The QET circuits for a degree-$d$ real polynomial without definite parity:
					(a) For even $d$. The angle $\vec{\phi}_{j}$ is defined as 
          $\vec{\phi}_{j}=(\phi_{j}^{(\mathrm{even})},\phi_{j-1}^{(\mathrm{odd})})$
					for $j=1,2,\dots,d$.
					(b) For odd $d$. The angle $\vec{\phi}_{j}$ is defined as 
					$\vec{\phi}_{j}=(\phi_{j-1}^{(\mathrm{even})},\phi_{j}^{(\mathrm{odd})})$
					for $j=1,2,\dots,d$.
          }
	\label{fig:QSVT_circuit_for_general_polynomial}
\end{figure*}

\begin{figure}[t]
	\centering
	\includegraphics[width=\linewidth]{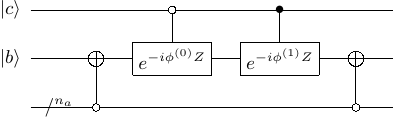}
	\caption{The gate $S_2(\vec{\phi})$. For $\vec{\phi}=(\phi^{(0)},\phi^{(1)}),b=\{0,1\},$ and $c=\{0,1\}$,
					 this gate is expressed as $S_2(\vec{\phi})=\sum_{b,c}\ket{cb}\bra{cb}\otimes e^{(-1)^b i\phi^{(c)}\Pi}$.}
	\label{fig:circuit_of_two_phi}
\end{figure}

We define a rectangular function in two forms: Open and closed rectangular functions.
For $\delta\in(0,1)$, an open rectangular function is defined by
\begin{equation}
  \mathrm{rect}(x) = \begin{cases}
    0 & |x|<\delta \\
    \frac{1}{2}  & |x|=\delta \\
    1  & \delta<|x|\leq1.
  \end{cases}
\end{equation}
This function is represented as a linear combination of two sign functions
\begin{equation}
  \mathrm{rect}(x) = 1+\frac{1}{2}\left(
    \mathrm{sgn}(x-\delta)+\mathrm{sgn}(-x-\delta)
  \right).
\end{equation}
Therefore, for $\Delta>0,\varepsilon\in(0, 1)$,
there exists a polynomial $P_{\delta, \Delta,\varepsilon}^{\mathrm{rect}}(x)$~\cite{martyn2021grand}
that satisfies the inequalities
\begin{align}
  \begin{split}
    \left|P_{\delta, \Delta,\varepsilon}^{\mathrm{rect}}(x)\right|&\leq\varepsilon,\quad x\in\left[-\delta+\frac{\Delta}{2},\delta-\frac{\Delta}{2}\right],\\
    \left|1-P_{\delta, \Delta,\varepsilon}^{\mathrm{rect}}(x)\right|&\leq\varepsilon,\\ 
        &x\in\left[-1,-\delta-\frac{\Delta}{2}\right]\cup\left[\delta+\frac{\Delta}{2}, 1\right],\\
    \left|P_{\delta, \Delta,\varepsilon}^{\mathrm{rect}}(x)\right|&\leq1,\quad x\in[-1,1]
  \end{split}
\end{align}
with the degree
\begin{equation} \label{eq:degree_of_poly_rect}
  d_{\mathrm{rect}}(\delta,\Delta,\varepsilon)
		=d_{\mathrm{sgn}}\left(\delta, \Delta, \frac{\varepsilon}{2}\right) - 1 
\end{equation}

On the other hand, a closed rectangular function is defined by
\begin{equation}
  \mathrm{rect}(x) = \begin{cases}
    1 & |x|<\delta \\
    \frac{1}{2}  & |x|=\delta \\
    0  & \delta<|x|\leq1.
  \end{cases}
\end{equation}
This function is also represented as a linear combination of two sign functions
\begin{equation}
  \mathrm{rect}(x) = \frac{1}{2}\left(
    \mathrm{sgn}(x+\delta)+\mathrm{sgn}(-x+\delta)
  \right).
\end{equation}
Thus, there exists a polynomial that $\varepsilon$ approximates the closed rectangular function with the degree in Eq.~\eqref{eq:degree_of_poly_rect}.

\subsubsection{General polynomial}

In specific problems, we encounter polynomials that do not satisfy conditions (i) and (ii).
This indicates a lack of definite parity and the maximum absolute value within the domain $[-1,1]$ exceeding 1. 
Such polynomials were discussed in~\cite{dong2021efficient}.
Here, we explicitly describe implementing a general polynomial through the QET.

Let $U_A$ be a $(\alpha,n_a,0)$ block encoding of $A$ and $P\in\mathbb{R}[x]$ be a real polynomial 
$P(x)=\sum_{j=0}^{d}c_j x^j$. We divide it into even and odd polynomials
\begin{align}
	\begin{split}
		P_{d}^{\mathrm{even}}(x) &= \sum_{l=0}^{\left\lceil \frac{d-1}{2}\right\rceil}c_{2l}x^{2l}, \\
		P_{d}^{\mathrm{odd}}(x) &= \sum_{l=1}^{\left\lceil \frac{d}{2}\right\rceil}c_{2l-1}x^{2l-1}.
	\end{split}
\end{align}
To ensure that each polynomial satisfies condition (ii), 
we normalize them using the maximum absolute values
\begin{align}
	\begin{split}
		C_{\max, d}^{\mathrm{even}} &= \max_{x\in[-1,1]}\left| P_{d}^{\mathrm{even}}(x)\right|, \\
		C_{\max, d}^{\mathrm{odd}} &= \max_{x\in[-1,1]}\left| P_{d}^{\mathrm{odd}}(x)\right|. 
	\end{split}
\end{align}
The normalized constants must be the same because we create a linear combination
of the normalized polynomials using two Hadamard gates; thus, we select the larger one
\begin{equation}
  C_{\max, d} = \max\left\{C_{\max, d}^{\mathrm{even}}, C_{\max, d}^{\mathrm{odd}}\right\}.
\end{equation}

Since the polynomials normalized by this constant meet both conditions, 
we can compute the phase factors $\Phi^{(\mathrm{even})}$ and 
$\Phi^{(\mathrm{odd})}$ from 
the corresponding even and odd polynomials. Figure~\ref{fig:QSVT_circuit_for_general_polynomial}
shows the QET circuits for a real polynomial without definite parity, representing
$(2C_{\max, d}, n_a+2, 0)$ block encoding of $P(A/\alpha)$.
These circuits construct the block encodings of the linear combination of the normalized even and odd polynomials
\begin{equation}
	\frac{\frac{1}{C_{\max,d}}\left[P_{d}^{\mathrm{even}}\left(\frac{A}{\alpha}\right)+P_{d}^{\mathrm{odd}}\left(\frac{A}{\alpha}\right)\right]}{2}
		= \frac{P\left(\frac{A}{\alpha}\right)}{2C_{\max,d}}
\end{equation}
If $P$ is a polynomial that $\varepsilon$ approximates a given function $f$ for $x\in[-1,1]$, then
the circuit becomes $(2C_{\max, d}, n_a+2, \varepsilon)$ block encoding of $f(A/\alpha)$.
The QET circuits have the gate $S_2(\vec{\phi})$ depicted in Fig.~\ref{fig:circuit_of_two_phi},
which has two angles of the phase factors for even and odd polynomials.
The number of queries to $U_A$ also equals the polynomial degree $d$.

\subsection{QLSA using the direct QSVT} \label{subsec:QLSA_by_direct_QSVT}
We present a review of the QLSA using the direct QSVT~\cite{martyn2021grand} to numerically compare the polynomial degrees 
of this algorithm and our algorithm in Sec.~\ref{sec:numerical_results}.
Let $A$ be a positive definite Hermitian matrix such that $1\leq\|A\|\leq\alpha$ ($\|A\|\leq1$ in~\cite{martyn2021grand}).
Note that the results below can be readily generalized to general matrices; 
hence, we adopt the term QSVT.
Let $\lambda_{\max}$ and $\lambda_{\min}$ be the maximum and minimum eigenvalues of $A$; then 
the condition number of $A$ is $\kappa=\|A\|\|A^{-1}\|=\lambda_{\max}/\lambda_{\min}\geq 1$. Since
$1\leq\lambda_{\max}=\| A\|\leq\alpha$, $1/\kappa\leq\lambda_{\min}$.
Therefore, the eigenvalues of $A/\alpha$ lie in the range $[\frac{1}{\kappa\alpha}, 1]$.
Here we introduce a domain denoted by $D_{\eta}=[-1,-1/\eta]\cup[1/\eta,1]$ for $\eta>0$.

The inverse matrix is represented as $A^{-1}=\sum_{j}\frac{1}{\lambda_j}\ket{\lambda_j}\bra{\lambda_j}$, where $\{\ket{\lambda_j}\}$ is the eigenvectors of $A$.
It can be rewritten as
$A^{-1}=g(A)$, where $g(x)=\frac{1}{x}$. The magnitude of a polynomial
is bounded by 1 in the QSVT framework.
Thus, the goal is to construct a polynomial approximating a function $f(x)=\frac{1}{2\kappa\alpha x}$
for $x\in D_{\kappa\alpha}$. This polynomial comprises two polynomial approximations:
$1/x$ and a rectangular function. 
As discussed in ~\cite{childs2017quantum,martyn2021grand}, for $0<\varepsilon<1/2$, 
the following odd polynomial $\varepsilon$ approximates $1/x$ for $D_{\kappa\alpha}$:
\begin{align} \label{eq:approximate_polynomial_2}
  &P_{\kappa, \alpha,\varepsilon}^{1/x}(x) \notag \\
  &= 4\sum_{j=0}^{(d_\mathrm{inv}-1)/2}(-1)^j\left[
    \frac{\sum_{i=j+1}^{b}\binom{2b}{b+i}}{2^{2b}}
  \right]T_{2j+1}(x),
\end{align}
where $T_{j}(x)$ is the $j$ th Chebyshev polynomial of the first kind, and
$b$ and $d_{\mathrm{inv}}$ are 
\begin{equation}
    b(\kappa, \alpha, \varepsilon)
		=\left\lceil(\kappa\alpha)^2\log\left(\frac{2\kappa\alpha}{\varepsilon}\right) \right\rceil,
\end{equation}
\begin{align}
  &d_{\mathrm{inv}}(\kappa, \alpha, \varepsilon) \notag \\
		&=2\left\lceil \frac{1}{2}\sqrt{b(\kappa, \alpha, \varepsilon)\log\left(\frac{8b(\kappa, \alpha, \varepsilon)}{\varepsilon}\right)} \right\rceil+1 \notag \\
    &= O\left(\kappa\alpha\log\left(\frac{\kappa\alpha}{\varepsilon}\right)\right).
\end{align}
Thus, $\frac{1}{2\kappa\alpha}P_{2\kappa, \alpha,\varepsilon}^{1/x}(x)$ is a polynomial
$\frac{\varepsilon}{2\kappa\alpha}$ approximating $f(x)$ for $x\in D_{2\kappa\alpha}$.

The absolute value of the above polynomial can be greater than 1 for $x\in[\frac{-1}{2\kappa\alpha},\frac{1}{2\kappa\alpha}]$.
Therefore, to ensure that the absolute value is bounded by 1, we must use the even function
close to 1 for $x\in D_{\kappa\alpha}$ and close to 0 for $x\in[\frac{-1}{2\kappa\alpha},\frac{1}{2\kappa\alpha}]$.
The domain
$\left[\frac{-1}{\kappa\alpha},\frac{-1}{2\kappa\alpha}\right] \cup \left[\frac{1}{2\kappa\alpha},\frac{1}{\kappa\alpha}\right]$
is a transition region between these two regions. 
A function that satisfies the above requirements is the approximate polynomial of
the open rectangular function for
$\delta=\frac{3}{4\kappa\alpha},\Delta=\frac{1}{2\kappa\alpha}$ in Sec.~\ref{subsubsec:approximate_polynomials_of_sign_and_rectangular_functions}.
This polynomial satisfies the inequalities
\begin{align}
	\begin{split}
		\left|P_{\frac{3}{4\kappa\alpha},\frac{1}{2\kappa\alpha},\varepsilon}^{\mathrm{rect}}(x)\right|&\leq\varepsilon,\quad x\in\left[-\frac{1}{2\kappa\alpha},\frac{1}{2\kappa\alpha}\right],\\
    \left|1-P_{\frac{3}{4\kappa\alpha},\frac{1}{2\kappa\alpha},\varepsilon}^{\mathrm{rect}}(x)\right|&\leq\varepsilon,\quad x\in D_{\kappa\alpha},\\
    \left|P_{\frac{3}{4\kappa\alpha},\frac{1}{2\kappa\alpha},\varepsilon}^{\mathrm{rect}}(x)\right|&\leq1,\quad x\in[-1,1]
	\end{split}
\end{align}
and the degree is 
\begin{equation}
	d_{\mathrm{rect}}\left(\frac{3}{4\kappa\alpha}, \frac{1}{2\kappa\alpha}, \varepsilon\right)
		= O\left(\kappa\alpha\log\left(1/\varepsilon\right)\right). \label{eq:degree_of_rect}
\end{equation}

We build the following target polynomial by multiplying those two polynomials:
\begin{equation}
  P_{\kappa,\alpha,\varepsilon}^{\mathrm{MI}}(x)
    \equiv \frac{1}{2\kappa\alpha}P_{2\kappa, \alpha, \frac{\varepsilon}{2}}^{1/x}(x)P_{\kappa,\alpha,\varepsilon'}^{\mathrm{rect}}(x),
\end{equation}
where 
\begin{equation}
	\varepsilon'
		=\min\left(\frac{2\varepsilon}{5\kappa\alpha},\frac{\kappa\alpha}{2d_{\mathrm{inv}}(2\kappa, \alpha, \varepsilon/2)}\right)
		=O\left(\frac{\varepsilon}{\kappa\alpha}\right).
\end{equation}
This polynomial satisfies $|P_{\kappa,\alpha,\varepsilon}^{\mathrm{MI}}(x)|\leq1$ 
for $x\in [-1,1]$ and $\frac{\varepsilon}{2\kappa\alpha}$ approximates $f(x)$
for $x\in D_{\kappa\alpha}$:
\begin{equation}
  \left|
    f(x) - P_{\kappa,\alpha,\varepsilon}^{\mathrm{MI}}(x)
  \right|\leq \frac{\varepsilon}{2\kappa\alpha}.
\end{equation}
The polynomial degree is the sum of the degrees of the two polynomials
\begin{align}
	d_{\mathrm{MI}}(\kappa, \alpha, \varepsilon) 
		&= d_{\mathrm{inv}}\left(2\kappa, \alpha, \frac{\varepsilon}{2}\right)
				+ d_{\mathrm{rect}}\left(\frac{3}{4\kappa\alpha},\frac{1}{2\kappa\alpha}, \varepsilon'\right) \notag \\
		&= O\left(\kappa\alpha\log\left(\frac{\kappa\alpha}{\varepsilon}\right)\right). \label{eq:degree_of_MI}
\end{align}

\subsection{Conjugate gradient method} \label{subsec:CG_method}
\begin{algorithm}[tbp]
	\caption{Conjugate gradient method.} \label{algorithm:conjugate_gradient_method}
	\begin{algorithmic}[1]
		\renewcommand{\algorithmicensure}{\textbf{Procedure:}}
		\ENSURE
		\STATE $\vec{r}_0=\vec{b}-A\vec{x}_0, \vec{p}_0=\vec{r}_0$
		\FOR {$k=0,1,\ldots$}
			\STATE $\alpha_k=\frac{\vec{r}_k\cdot\vec{r}_k}{\vec{p}_k\cdot A\vec{p}_k}$
			\STATE $\vec{x}_{k+1}=\vec{x}_k+\alpha_k \vec{p}_k$
			\STATE $\vec{r}_{k+1}=\vec{r}_k-\alpha_k A \vec{p}_k$
			\IF {$\left\| \vec{r}_{k+1}\right\| \leq \varepsilon\|\vec{b}\|$}
				\STATE{STOP}
			\ENDIF
			\STATE $\beta_k=\frac{\vec{r}_{k+1}\cdot\vec{r}_{k+1}}{\vec{r}_k\cdot\vec{r}_k}$
			\STATE $\vec{p}_{k+1}=\vec{r}_{k+1}+\beta_k \vec{p}_k$
		\ENDFOR
	\end{algorithmic}
\end{algorithm}

The conjugate gradient method~\cite{shewchuk1994introduction,saad2003iterative} 
stands as a classical algorithm for solving
a linear system of equations for a positive definite Hermitian matrix
$A\in\mathbb{R}^{N\times N}$.
This method belongs to a category of iterative techniques known as the Krylov subspace methods.
The Krylov subspace is formed by vectors 
$\vec{v}, A\vec{v},A^2\vec{v},\dots,A^{k-1}\vec{v} (k\leq N)$.
Here $\vec{v}$ denotes a given $N$-dimensional vector.
The vectors in the Krylov subspace are expressed as the product of
the polynomial $P(A)$ and the initial vector $\vec{v}$. Therefore, 
the vectors of the CG method are also represented as the same product. 

Algorithm~\ref{algorithm:conjugate_gradient_method} outlines  the steps of the CG method.
In each iteration, three vectors are generated: An approximate solution vector
$\vec{x}_k$, a residual vector $\vec{r}_k = \vec{b}-A\vec{x}_k$, and
a search vector $\vec{p}_k$. When $\vec{x}_0 = 0$, these vectors can be expressed
as the product of the polynomial $P(A)$ and the initial vector $\vec{b}$ as follows:
\begin{align}
	\begin{split}
		\vec{x}_k &= P_{x, k}(A)\vec{b} = \sum_{l=0}^{k-1}x_{k,l}A^l\vec{b}, \\
		\vec{r}_k &= P_{r, k}(A)\vec{b} = \sum_{l=0}^{k}r_{k,l}A^l\vec{b}, \\
		\vec{p}_k &= P_{p, k}(A)\vec{b} = \sum_{l=0}^{k}p_{k,l}A^l\vec{b}.
	\end{split}
\end{align}
Based on Algorithm~\ref{algorithm:conjugate_gradient_method}, the coefficients of 
those three polynomials are updated through the following equations: 
For $l=0,1,\dots,k+1$,
\begin{align} \label{eq:update_coefficients} 
	\begin{split}
		x_{k+1,l} &= x_{k,l}+\alpha_k p_{k,l}, \\
		r_{k+1,l} &= r_{k,l}-\alpha_k p_{k,l-1}, \\
		p_{k+1,l} &= r_{k+1,l}+\beta_k p_{k,l},
	\end{split}
\end{align}
where $p_{k,-1}=x_{k,k}=x_{k,k+1}=r_{k,k+1}=p_{k,k+1}=0$. 
In each iteration, $\vec{x}_k$, $\vec{r}_k$, and $\vec{p}_k$ are generated
by constructing the polynomials using the coefficients determined by 
the above equations.

\begin{figure}[tbp]
	\centering
	\begin{minipage}[c]{\linewidth}
		\centering
    \includegraphics[width=0.9\linewidth]{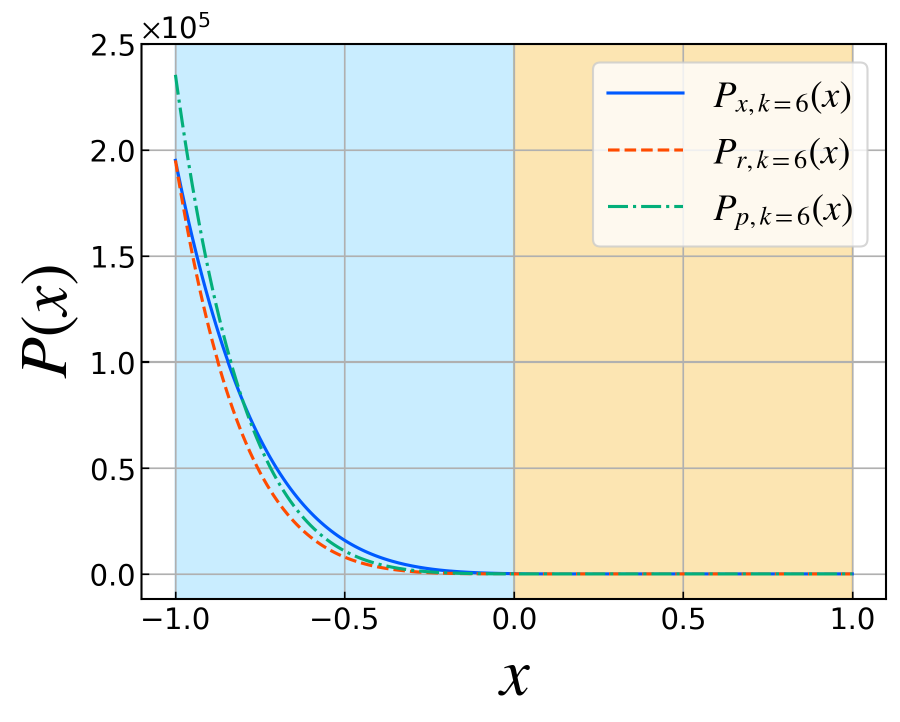}
		\put(-210,155){\Large{(a)}}
		\label{fig:polynomials_-1_1}
	\end{minipage} \\ 
	\begin{minipage}[c]{\linewidth}
		\centering
    \includegraphics[width=0.9\linewidth]{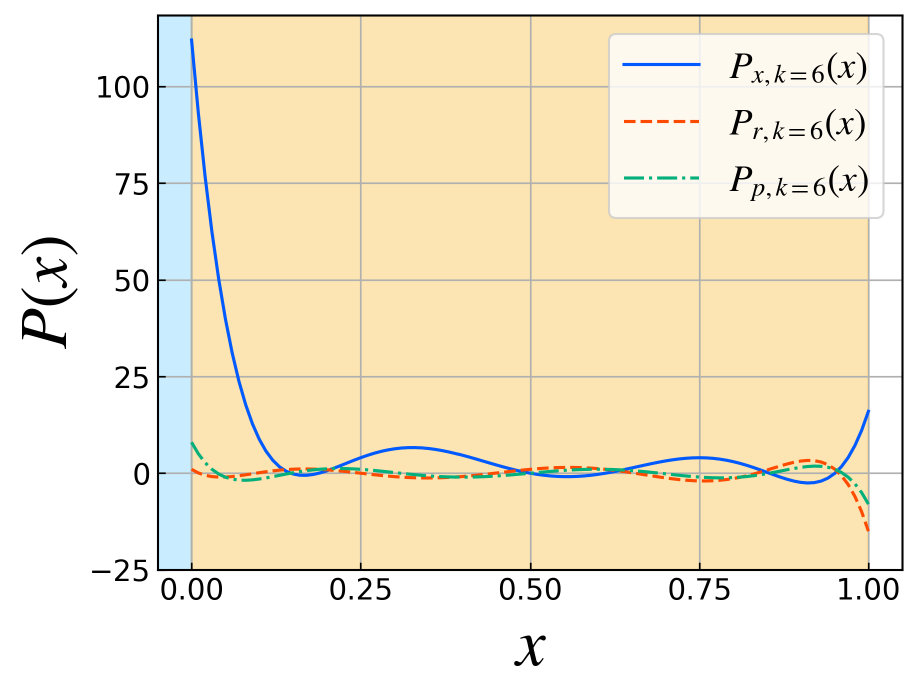}
		\put(-210,150){\Large{(b)}}
		\label{fig:polynomials_0_1}
	\end{minipage}
  \caption{Plots of polynomials obtained by applying the CG method 
					 to the one-dimensional Poisson equation: (a) For $x\in[-1,1]$ and
					 (b) for $x\in[0,1]$.
					 The orange area represents the positive side ($x\in[0,1]$),
					 and the blue area represents the negative side ($x\in[-1,0]$).}
  \label{fig:polynomials}
\end{figure}

We address a potential problem concerning polynomials in the CG method. 
Figure~\ref{fig:polynomials} shows
the polynomials obtained by applying this method to the one-dimensional
Poisson equation (refer to Sec.~\ref{sec:numerical_results} for more details). 
Figure~\ref{fig:polynomials}(a) highlights the significant scaling of these polynomials
for $x\in[-1,1]$, whereas Fig.~\ref{fig:polynomials}(b) shows 
the relatively small scaling for $x\in[0,1]$.
This enormous value arises because the polynomial coefficients are amplified due to $\alpha_k>1$ as
iterations progress.
The maximum absolute values on the positive side ($x\in[0,1]$) are smaller because
the effects of the powers of the eigenvalues whose magnitudes are less than one and the huge coefficients cancel each other out.
As explained in Sec.~\ref{subsec:QET}, the QET framework necessitates the normalization of
polynomials by their maximum absolute values.
Normalizing the polynomials in Fig.~\ref{fig:polynomials}(a) using their maximum absolute values 
results in an extremely small success probability of the algorithm, 
indicating a significant increase in the query complexity.
In Sec.~\ref{subsec:positive_side_QET} we provide a technique to address and overcome this challenge.

\subsection{Swap test}  \label{subsec:swap_test}
\begin{figure}[tbp]
	\centering
	\includegraphics[width=\linewidth]{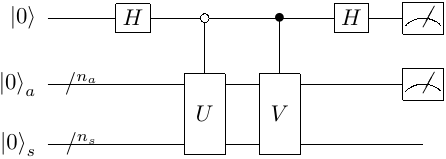}
  \caption{Circuit of the swap test for block encodings.}
  \label{fig:circuit_of_SWAP_test_for_block_encoding}
\end{figure}

We explain a method for computing the inner product of states multiplied by
block encodings. This method is a slightly modified version of the swap test~\cite{buhrman2001quantum}. 
Let $A$ and $B$ acting as $A\ket{0}_s=\ket{\psi}_s$ and $B\ket{0}_s=\ket{\phi}_s$
(these matrices' norm is less than or equal to 1);
Then ${}_s\langle \psi|\phi\rangle_s$ is an inner product to be estimated.
Let $U$ and $V$ be $(1,n_a,0)$ block encodings
of $A$ and $B$ acting on the same qubits as follows:
\begin{align}
	\begin{split}
		U\ket{0}_a\ket{0}_s &= \ket{0}_a A\ket{0}_s + \ket{\bot}  = \ket{0}_a \ket{\psi}_s + \ket{\bot}, \\ 
		V\ket{0}_a\ket{0}_s &= \ket{0}_a B\ket{0}_s + \ket{\bot'} = \ket{0}_a \ket{\phi}_s + \ket{\bot'},
	\end{split}
\end{align}
where $\ket{\bot}$ and $\ket{\bot'}$ are states orthogonal to $\ket{0}_a$.
In the circuit shown in Fig.~\ref{fig:circuit_of_SWAP_test_for_block_encoding},
the state before measurement is
\begin{equation}
  \frac{1}{2}\ket{0}\ket{0}_a(\ket{\psi}_s+\ket{\phi}_s)
    + \frac{1}{2}\ket{1}\ket{0}_a(\ket{\psi}_s-\ket{\phi}_s) + \ket{\bot''},
\end{equation}
where $\ket{\bot''}$ is a state orthogonal to $\ket{0}\ket{0}_a$ and $\ket{1}\ket{0}_a$.
The probabilities of obtaining $\ket{0}\ket{0}_a$ and $\ket{1}\ket{0}_a$, denoted by 
$p_0$ and $p_1$, are 
\begin{align}
	\begin{split}
		p_0 &= \frac{{}_s\langle \psi|\psi\rangle_s+{}_s\langle \phi|\phi\rangle_s+2\mathrm{Re}({}_s\langle \psi|\phi\rangle_s)}{4}, \\
		p_1 &= \frac{{}_s\langle \psi|\psi\rangle_s+{}_s\langle \phi|\phi\rangle_s-2\mathrm{Re}({}_s\langle \psi|\phi\rangle_s)}{4}. 
	\end{split}
\end{align}
Therefore, the real part of the inner product 
is expressed by
\begin{equation}
  \mathrm{Re}({}_s\langle \psi|\phi\rangle_s) = p_0 - p_1.
\end{equation}

To achieve the estimate with $\varepsilon$ precision, $O(1/\varepsilon^2)$ samples from the circuit Fig.~\ref{fig:circuit_of_SWAP_test_for_block_encoding} are required. Obviously, quantum amplitude estimation~\cite{brassard2002quantum,suzuki2020amplitude,grinko2021iterative}
reduces the total queries to $U$ and $V$ to $O(1/\varepsilon)$. However, this paper avoids
this algorithm to maintain circuits shallow.

\section{Proposed methods} \label{sec:proposed_methods}
\subsection{Positive-side QET} \label{subsec:positive_side_QET}
\begin{figure}[tbp]
	\centering
	\begin{minipage}[c]{\linewidth}
		\centering
    \includegraphics[width=0.9\linewidth]{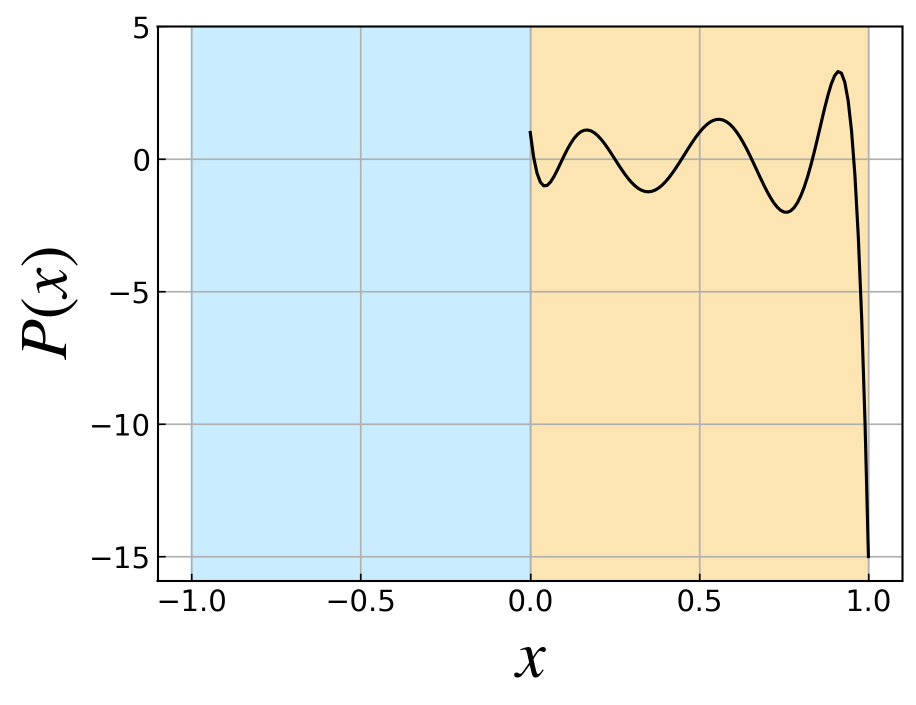}
		\put(-210,155){\Large{(a)}}
		\label{fig:original_polynomial}
	\end{minipage} \\ 
	\begin{minipage}[c]{\linewidth}
		\centering
    \includegraphics[width=0.9\linewidth]{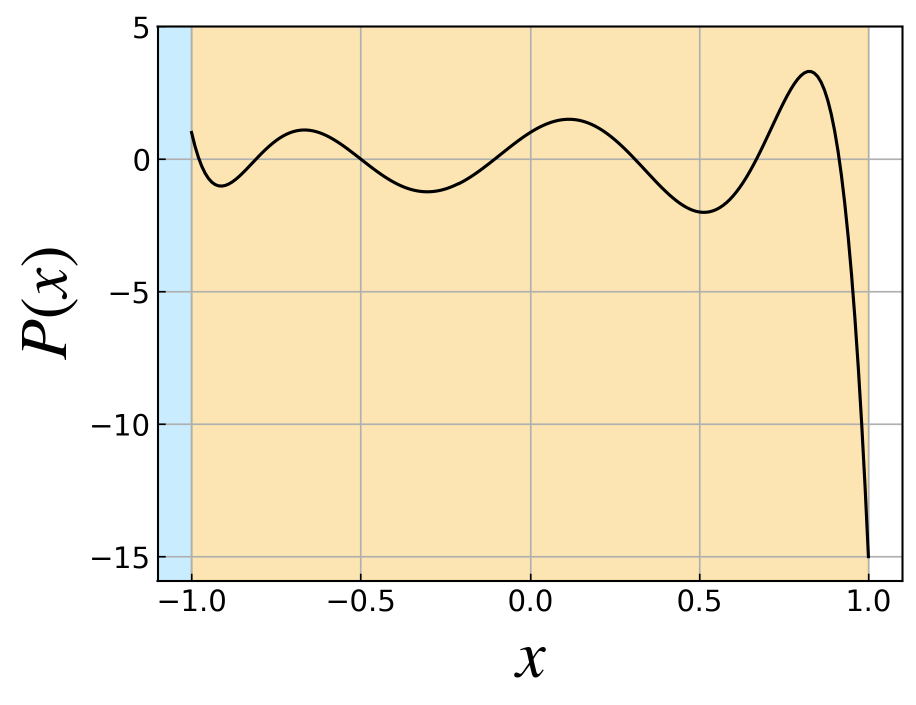}
		\put(-210,150){\Large{(b)}}
		\label{fig:shifted_and_enlarged_polynomial}
	\end{minipage}
  \caption{
		Illustration of shifting and enlarging a polynomial. The black curve represents $P_{r,k=6}(x)$ in Fig.~\ref{fig:polynomials}.
		The orange region represents the positive side ($x\in[0,1]$) of the original polynomial, 
		and the blue region represents the negative side ($x\in[-1,0]$).
		For clarity, we deliberately omit the negative side of the polynomial,
		which has a large magnitude.
	}
  \label{fig:positive_side_QET}
\end{figure}

We introduce a novel technique named positive-side QET to address
the issue discussed in Sec.~\ref{subsec:CG_method}.
Let $A$ be a positive semidefinite Hermitian matrix,
$U_A$ be a $(\alpha,n_a,0)$ block encoding of $A$, and $P$ be a target real polynomial. 
The eigenvalues of $A/\alpha$ exist in the range $[0,1]$,
indicating that the polynomial's relevant domain lies on the positive side ($x\in[0,1]$).
However, in the QET framework, the domain of interest is $[-1,1]$.
To utilize only the positive side of $P(x)$, we eliminate the negative side ($x\in[-1,0]$) 
from the domain $[-1,1]$ by shifting and enlarging this polynomial. 
Specifically, we shift it by $-1/2$ in the $x$-direction and
enlarge it by a factor of 2 in the $x$-direction. We denote the modified polynomial by
\begin{equation}
	P^{+}(x)=P\left(\frac{x+1}{2}\right).
\end{equation}
Figure~\ref{fig:positive_side_QET} visually represents these operations.

If we compute phase factors from the modified polynomial and use
the block encoding $U_A$, then we have
\begin{equation}
	P^+(A/\alpha)=P\left(\frac{A/\alpha+I}{2}\right)\neq P(A/\alpha),
\end{equation}
which does not match the target polynomial. Consequently, $U_A$ must be substituted with
the $(1,n_a,0)$ block encoding $U_{A'}$ of 
\begin{equation}
	A'=2A/\alpha-I.
\end{equation}
This adjustment leads to the construction of
the target polynomial
\begin{equation}
	P^+(A')=P\left(\frac{2A/\alpha-I+I}{2}\right)=P(A/\alpha).
\end{equation}
Therefore, this technique allows us to amplify the success probability by a factor of $\max_{x \in [-1,1]} |P(x)| / \max_{x \in [0,1]} |P(x)|$ at only the additional cost of constructing $U_{A'}$. 
The success probability can be improved dramatically if this ratio is large. 
Thus, for a positive semidefinite Hermitian matrix, a polynomial with a huge absolute value on the negative side can be handled more efficiently by using $U_{A'}$ and $P^{+}(x)$ than by naively using $U_{A}$ and $P(x)$.

We describe a procedure for constructing a degree-$d$ general polynomial by combining 
the positive-side QET with the method described in Sec.~\ref{subsec:QET}. 
First, we derive the polynomial $P^+(x)$
from the original polynomial $P(x)$ and divide it into 
even and odd polynomials. Then we normalize these polynomials using
the larger maximum absolute value for $x\in[0,1]$ instead of $x\in[-1,1]$,
denoted by $C_{\max, d}$, and
compute the corresponding phase factors. Finally, we construct a circuit
depicted in Fig.~\ref{fig:QSVT_circuit_for_general_polynomial}
using these phase factors and $d$ queries to $U_{A'}$. This circuit 
represents $(2C_{\max, d}, n_a+2, 0)$ block encoding of $P(A/\alpha)$.

Note that the above technique requires efficiently implementing the block encoding $U_{A'}$ of $A'$.
Importantly, if $A$ is sparse, then $A'$ is also sparse.
Therefore, we can efficiently create the block encoding of $A'$ for a sparse matrix $A$.
The details of this construction remain a challenging task; however, some proposed methods exist.
For instance, Refs.~\cite{camps2022explicit,sunderhauf2024block} discuss schemes for 
building explicit circuits for block encodings of structured sparse matrices.

Additionally, we propose explicit circuits for $U_{A'}$
from $U_A$, as detailed in Appendix~\ref{appendix:explicit_circuits_for_block_encoding_of_of_A_dash}.
We present two methods with and without a gap $\Lambda\in(0,\frac{\alpha}{2})$.
This gap is related to the $A$'s eigenvalues $\{\lambda_j\}$ such that
\begin{equation}
  \Lambda \leq \lambda_{j} \leq \alpha-\Lambda.
\end{equation}
The method with $\Lambda$
requires $O(\frac{\alpha}{\Lambda}\log(1/\varepsilon))$ additional queries to $U_A$, while
the method without it demands $O(\frac{1}{\varepsilon}\log(1/\varepsilon))$ additional queries.
Both methods involve the linear combination of unitaries and 
linear amplification~\cite{gilyen2019quantum,martyn2021grand,low2019hamiltonian}.
The method selection should be appropriate based on the magnitude of $\Lambda$, $\alpha$, and $\varepsilon$.

\subsection{Quantum conjugate gradient method using the quantum eigenvalue transformation} \label{subsec:QCG_method}
\begin{figure}[tbp]
	\centering
  \includegraphics[width=\linewidth]{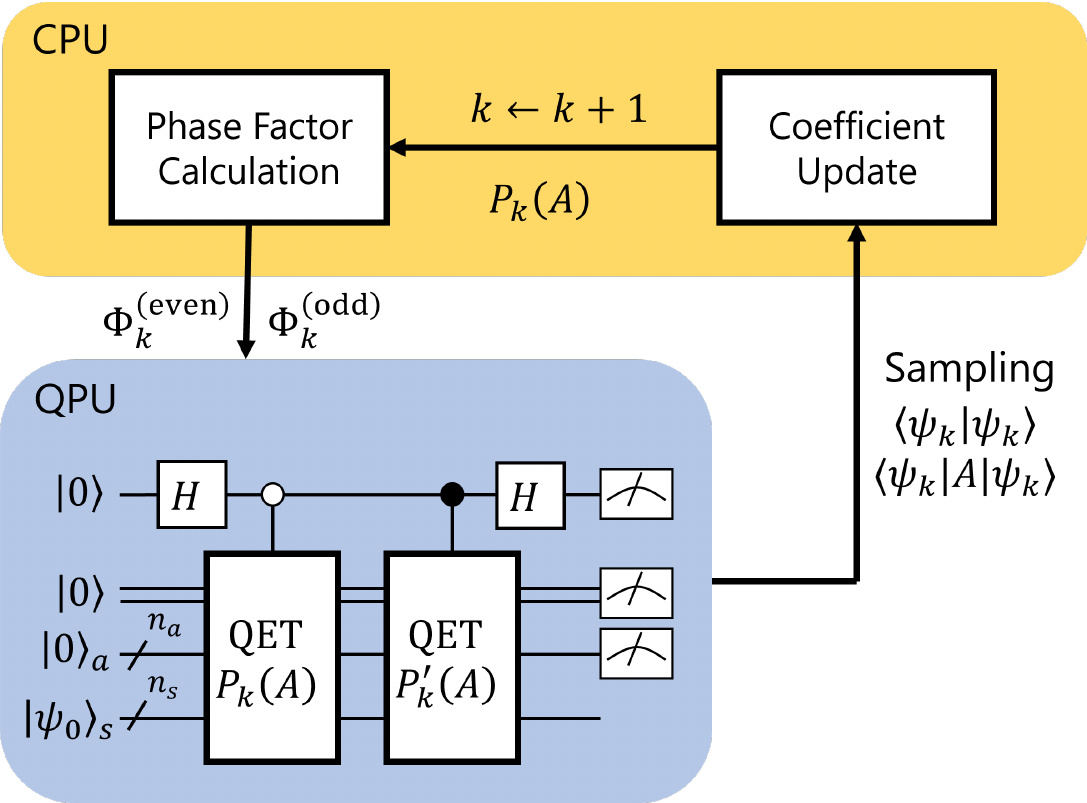}
  \caption{Illustration of the quantum conjugate gradient method using the 
					quantum eigenvalue transformation.}
  \label{fig:overview_of_QCG_using_QET}
\end{figure}

The QCG method using the QET is a hybrid quantum-classical algorithm. Quantum computations include
polynomial implementations through the QET and evaluations of inner products using the swap test described in
Sec.~\ref{subsec:swap_test}. Classical computations cover updating coefficients according to Eq.~\eqref{eq:update_coefficients}
and calculating the phase factors based on the updated coefficients. Figure~\ref{fig:overview_of_QCG_using_QET} illustrates 
the QCG method using the QET. 

\begin{algorithm*}[t]
	\caption{Quantum conjugate gradient method using the quantum eigenvalue transformation.} \label{algorithm:QCG_using_QET}
	\begin{algorithmic}[1]
		\renewcommand{\algorithmicensure}{\textbf{Procedure:}}
		\ENSURE
		\STATE $\ket{x_0}=0, \ket{r_0}=\ket{p_0}=\frac{\ket{b}}{\|\ket{b}\|}$
		\STATE Estimate $\braket{p_{0}|p'_{0}}$ using the swap test with $\delta$ precision.
		\FOR {$k=0,1,\ldots, m$}
			\STATE $\alpha_k=\frac{\braket{r_k|r_k}}{\braket{p_k|p'_k}}$
			\FOR {$l=0,1,\ldots,k,k+1$}
				\STATE $x_{k+1,l}=x_{k,l}+\alpha_k p_{k,l}$
				\STATE $r_{k+1,l}=r_{k,l}-\alpha_k p_{k,l-1}$
			\ENDFOR

			\STATE Compute $R_{\max, k+1}$ from $\{r_{k+1,l}\}$ and the phase factors for the normalized even and odd polynomials.

			\STATE Estimate $\frac{\braket{r_{k+1}|r_{k+1}}}{4R_{\max, k+1}^{2}}$ using the swap test
						 with $\frac{\delta}{4R_{\max, k+1}^{2}}$ precision.
			\IF {$\left\| \ket{r_{k+1}}\right\| \leq \frac{\|A\|\varepsilon}{\kappa\|\ket{b}\|}$}
				\STATE{BREAK}
			\ENDIF
			\STATE $\beta_{k}=\frac{\braket{r_{k+1}|r_{k+1}}}{\braket{r_{k}|r_{k}}}$
			\FOR {$l=0,1,\ldots,k,k+1$}
				\STATE $p_{k+1,l}=r_{k+1,l}+\beta_k p_{k,l}$
			\ENDFOR

			\STATE Compute $P_{\max, k+1}$ and $P'_{\max, k+1}$ from $\{p_{k+1,l}\}$ and the phase factors for the normalized even and odd polynomials.
			\STATE Estimate $\frac{\braket{p_{k+1}|p'_{k+1}}}{4P_{\max,k+1}P'_{\max,k+1}}$ using the swap test
						 with $\frac{\delta}{4P_{\max,k+1}P'_{\max,k+1}}$ precision. 
		\ENDFOR
		\STATE Compute $X_{\max, m+1}$ from $\{x_{m+1,l}\}$ and the phase factors for the normalized even and odd polynomials.
		\STATE Prepare the state proportional to $\frac{1}{2X_{\max, m+1}}\ket{x_{m+1}}$.
	\end{algorithmic}
\end{algorithm*}

We introduce some notations to elaborate on the QCG method in detail.
Although our goal is to solve $A\ket{x}=\ket{b}$, 
the $(\alpha,n_a,0)$ block encoding of $A$ and the initial vector 
$\frac{\ket{b}}{\|\ket{b}\|}$ are applied to the QCG method.
Consequently, the vectors are represented as
\begin{align}
    \begin{split}
        \ket{x_k} &= P_{x, k}\left(\frac{A}{\alpha}\right)\frac{\ket{b}}{\|\ket{b}\|} = \sum_{l=0}^{k-1}x_{k,l}\left(\frac{A}{\alpha}\right)^l\frac{\ket{b}}{\|\ket{b}\|}, \\
        \ket{r_k} &= P_{r, k}\left(\frac{A}{\alpha}\right)\frac{\ket{b}}{\|\ket{b}\|} = \sum_{l=0}^{k}r_{k,l}\left(\frac{A}{\alpha}\right)^l\frac{\ket{b}}{\|\ket{b}\|}, \\
        \ket{p_k} &= P_{p, k}\left(\frac{A}{\alpha}\right)\frac{\ket{b}}{\|\ket{b}\|} = \sum_{l=0}^{k}p_{k,l}\left(\frac{A}{\alpha}\right)^l\frac{\ket{b}}{\|\ket{b}\|}, \\
        \ket{p'_k} &= \frac{A}{\alpha}\ket{p_k} = \frac{A}{\alpha}P_{p, k}\left(\frac{A}{\alpha}\right)\frac{\ket{b}}{\|\ket{b}\|}.
    \end{split}
\end{align}
The approximate solution vector $\ket{x_k}$ approximates $\frac{\alpha}{\|\ket{b}\|}\ket{x}$, and
the residual vector is expressed as $\ket{r_k}=\frac{\ket{b}}{\|\ket{b}\|}-\frac{A}{\alpha}\ket{x_k}$.
The corresponding absolute maximum values of the above polynomials for $x\in[0,1]$
are denoted by $X_{\max, k}, R_{\max, k}, P_{\max, k}$, and $P'_{\max, k}$. 
The accuracy of inner product estimations is $\delta>0$. 
Algorithm~\ref{algorithm:QCG_using_QET} summarizes the QCG method using the QET. 

In classical computing, the convergence of the CG method has been extensively
examined~\cite{shewchuk1994introduction,saad2003iterative}. 
This result is applied to the QCG method, and the convergence is expressed through the inequality
\begin{align}
	&\left\|\frac{\alpha}{\|\ket{b}\|}\ket{x}-\ket{x_k}\right\|_{\frac{A}{\alpha}} \notag \\
		&\leq 2\left(
			\frac{\sqrt{\kappa}-1}{\sqrt{\kappa}+1}
		\right)^k\left\|\frac{\alpha}{\|\ket{b}\|}\ket{x}-\ket{x_0}\right\|_{\frac{A}{\alpha}}
\end{align}
where $\|\cdot\|_{A}$ represents the $A$-norm such that $\|\ket{\psi}\|_{A}=\sqrt{\bra{\psi}A\ket{\psi}}$.
Since $\ket{x_0}=0$ and 
$\|\ket{x}\|_{\frac{A}{\alpha}}=\|\sqrt{\frac{A}{\alpha}}\ket{x}\|\leq\|\ket{x}\|\leq\frac{\kappa\|\ket{b}\|}{\|A\|}$, we have
\begin{equation}
	\left\|\ket{x}-\frac{\|\ket{b}\|}{\alpha}\ket{x_k}\right\|_{\frac{A}{\alpha}}
		\leq2\left(
			\frac{\sqrt{\kappa}-1}{\sqrt{\kappa}+1}
		\right)^k\frac{\kappa\|\ket{b}\|}{\|A\|}.
\end{equation}
A sufficient number of iterations are executed to decrease the above norm 
by a factor of $\varepsilon>0$;
that is, $\left\|\ket{x}-\frac{\|\ket{b}\|}{\alpha}\ket{x_k}\right\|_{\frac{A}{\alpha}}\leq\varepsilon$. As a result,
the maximum number of iterations is given by
\begin{align} \label{eq:number_of_iterations}
	k &= \left\lceil
		\frac{1}{2} \sqrt{\kappa}\ln\left(
			\frac{2\kappa\|\ket{b}\|}{\|A\|\varepsilon}
		\right)
	\right\rceil \notag \\
 &= O\left(\sqrt{\kappa}\log\left(\frac{\kappa\|\ket{b}\|}{\|A\|\varepsilon}\right)\right).
\end{align}

However, since the solution $\ket{x}$ is unknown, we cannot access $\left\|\ket{x}-\frac{\|\ket{b}\|}{\alpha}\ket{x_k}\right\|_{\frac{A}{\alpha}}$.
Instead, we use the residual $\|\ket{r_k}\|$. If the inequality 
$\|\ket{r_k}\|\leq\frac{\|A\|\varepsilon}{\kappa\|\ket{b}\|}$ holds, then the following inequalities also hold:
\begin{align}
	\left\|\ket{x}-\frac{\|\ket{b}\|}{\alpha}\ket{x_k}\right\|
		&=\left\|A^{-1}\ket{b}-\frac{\|\ket{b}\|}{\alpha}A^{-1}A\ket{x_k}\right\| \notag \\
		&\leq\|A^{-1}\|\|\ket{b}\|\|\ket{r_k}\| \notag \\
		&\leq\|A^{-1}\|\|\ket{b}\|\cdot\frac{\|A\|\varepsilon}{\kappa\|\ket{b}\|}=\varepsilon, 
\end{align}
and
\begin{align}
	\left\|\ket{x}-\frac{\|\ket{b}\|}{\alpha}\ket{x_k}\right\|_{\frac{A}{\alpha}}
		&=\left\|\sqrt{\frac{A}{\alpha}}\left(\ket{x}-\frac{\|\ket{b}\|}{\alpha}\ket{x_k}\right)\right\| \notag \\
		&\leq\left\|\sqrt{\frac{A}{\alpha}}\right\|\left\|\ket{x}-\frac{\|\ket{b}\|}{\alpha}\ket{x_k}\right\| \notag \\
		&\leq\varepsilon.
\end{align}
If we define $\|\ket{r_k}\|\leq\frac{\|A\|\varepsilon}{\kappa\|\ket{b}\|}$ as a convergence criterion, 
the algorithm stops after the number of iterations in Eq.~\eqref{eq:number_of_iterations}.

In practical application, the computation does not precisely converge in the above number of iterations
due to estimate errors from swap tests. Therefore, we need sufficiently accurate estimations.
According to the convergence criterion, the inequality $\braket{r_k|r_k}\leq\left(\frac{\|A\|\varepsilon}{\kappa\|\ket{b}\|}\right)^2$ 
holds; thus, we require $\delta<\left(\frac{\|A\|\varepsilon}{\kappa\|\ket{b}\|}\right)^2$.

Under this assumption, the computation of the QCG method using the QET can also stop
after $m=O\left(\sqrt{\kappa}\log\left(\frac{\kappa\|\ket{b}\|}{\|A\|\varepsilon}\right)\right)$ iterations. 
Since the circuit for estimating $\braket{r_{m+1}|r_{m+1}}$ exhibits the highest depth
in Algorithm~\ref{algorithm:QCG_using_QET}, 
the maximum circuit depth is
\begin{equation}
	2(m+1) = O\left(\sqrt{\kappa}\log\left(\frac{\kappa\|\ket{b}\|}{\|A\|\varepsilon}\right)\right).
\end{equation}
This outcome indicates that this algorithm improves the maximum circuit depth compared to 
other QLSAs~\cite{ambainis2010variable,
childs2017quantum,wossnig2018quantum,shao2018quantum,
gilyen2019quantum,lin2020optimal,martyn2021grand}.

Here we discuss the total query complexity of the QCG method using the QET.
In the $k$ th iteration, the total number of queries to the block encoding scales as
\begin{align}
	\begin{split}
		Q_k\equiv O\Bigg(
			&\frac{32(k+1)R_{\max, k+1}^{4}}{\delta^2} \\
			&\quad +\frac{16(2k+3)P_{\max, k+1}^{2}P_{\max, k+1}^{'2}}{\delta^2}
			\Bigg).
	\end{split}
\end{align}
Generating $\frac{1}{2X_{\max, m+1}}\ket{x_{m+1}}$ requires
$m$ queries, and obtaining the state $\frac{\ket{x_{m+1}}}{\|\ket{x_{m+1}}\|}$ 
demands $O\left(\frac{2X_{\max, m+1}^{2}}{\|\ket{x_{m+1}}\|^2}\right)$ circuit runs.
Therefore, the total query complexity is expressed as
\begin{equation} 
	O\left(
    \sum_{k=0}^{m}Q_k + m\left(\frac{2X_{\max, m+1}}{\|\ket{x_{m+1}}\|}\right)^2
  \right). \label{eq:total_query_complexity_of_QCG_by_QET}
\end{equation}

For simplicity, we assume that the maximum absolute values of the polynomials depend on
the iteration number $k$ as follows:
\begin{align}
	\begin{split} \label{eq:dependencies_of_max}
		X_{\max, k}&=O(k^{\gamma_x}), \quad R_{\max, k}=O(k^{\gamma_r}), \\
		P_{\max, k}&=O(k^{\gamma_p}), \quad P'_{\max, k}=O(k^{\gamma_{p'}}),
	\end{split}
\end{align}
where $\gamma_{r},\gamma_{p},\gamma_{p'},\gamma_{x}\geq0$.
Consequently, Eq.~\eqref{eq:total_query_complexity_of_QCG_by_QET} scales as
\begin{align}
	&O\left(
    \sum_{k=0}^{m+1}\frac{k(k^{4\gamma_{r}}+k^{2(\gamma_{p}+\gamma_{p'})})}{\delta^2}+ \frac{m^{1+2\gamma_x}}{\|\ket{x_{m+1}}\|^2}
  \right) \notag \\
		&= O\left(
			\frac{m^{2}(m^{4\gamma_{r}}+m^{2(\gamma_{p}+\gamma_{p'})})}{\delta^2}+ \frac{m^{1+2\gamma_x}}{\|\ket{x_{m+1}}\|^2}
		\right) \notag \\
		&= O\left(
			\left[\frac{\kappa^{1+\gamma}}{\delta^2}+  \frac{\kappa^{\frac{1}{2}+\gamma_x}}{\|\ket{x_{m+1}}\|^2}\right]\mathrm{polylog}\left(\frac{\kappa\|\ket{b}\|}{\|A\|\varepsilon}\right)
		\right), \label{eq:total_query_complexity_of_QCG_by_QET2}
\end{align}
where in the last equality, we used $\gamma\equiv\max\{2\gamma_r, \gamma_{p}+\gamma_{p'}\}$ 
and $m=O\left(\sqrt{\kappa}\log\left(\frac{\kappa\|\ket{b}\|}{\|A\|\varepsilon}\right)\right)$. 
Additionally, we assume that the estimation accuracy is such that
$c\left(\frac{\|A\|\varepsilon}{\kappa\|\ket{b}\|}\right)^2<\delta<\left(\frac{\|A\|\varepsilon}{\kappa\|\ket{b}\|}\right)^2$ for a constant $c\in(0,1)$.
The approximate solution is assumed to satisfy $\|\ket{x_k}\| = \Omega(1)$ as a worst-case.
Equation~\eqref{eq:total_query_complexity_of_QCG_by_QET2} becomes
\begin{equation}
	O\left(
		\left[\kappa^{5+\gamma}\left(\frac{\|\ket{b}\|}{\|A\|\varepsilon}\right)^4+ 
        \kappa^{\frac{1}{2}+\gamma_x}\right]\mathrm{polylog}\left(\frac{\kappa\|\ket{b}\|}{\|A\|\varepsilon}\right)
	\right).
\end{equation}
Note that this scaling does not include the subnormalization factor $\alpha$.
This absence occurs when the block encoding of $A'=2\frac{A}{\alpha}-I$ is present or is constructed using the method without the gap of Appendix~\ref{appendix:explicit_circuits_for_block_encoding_of_of_A_dash}.
However, if the method with the gap is utilized,
the total query complexity depends on $\alpha$.

The validity of the assumption in Eq.~\eqref{eq:dependencies_of_max} varies; 
it holds in some cases but not in others, as demonstrated by the numerical experiments 
in Sec.~\ref{sec:numerical_results}. 
In a particular simulation, the maximum absolute values depend on $\kappa$ instead of 
$k$ as follows:
\begin{align}
	\begin{split} \label{eq:dependencies_of_max_2}
		X_{\max, k}&=O(\kappa^{\eta_x}), \quad R_{\max, k}=O(\kappa^{\eta_r}), \\
		P_{\max, k}&=O(\kappa^{\eta_p}), \quad P'_{\max, k}=O(\kappa^{\eta_{p'}}),
	\end{split}
\end{align}
where $\eta_{r},\eta_{p},\eta_{p'},\eta_{x}\geq0$.
Accordingly, the total query complexity scales as
\begin{equation}
	O\left(
		\left[\kappa^{5+\eta}\left(\frac{\|\ket{b}\|}{\|A\|\varepsilon}\right)^4
        +\kappa^{\frac{1}{2}+2\eta_x}\right]\mathrm{polylog}\left(\frac{\kappa\|\ket{b}\|}{\|A\|\varepsilon}\right)
	\right),
\end{equation}
where $\eta\equiv\max\{4\eta_r, 2(\eta_{p}+\eta_{p'})\}$.

In either case,
the scaling for $\kappa$ and $\varepsilon$ is less favorable than that of 
the other QLSAs~\cite{ambainis2010variable,
childs2017quantum,wossnig2018quantum,shao2018quantum,
gilyen2019quantum,lin2020optimal,martyn2021grand}.
This scaling results from numerous circuit runs to attain precise estimates of 
inner products through the swap tests.
Our algorithm can be interpreted as prioritizing a low circuit depth at the expense of 
an increased number of circuit runs.

It is essential to highlight that the total query complexity is worst-case
because the precision required to estimate all inner products must be
less than $\left(\frac{\|A\|\varepsilon}{\kappa\|\ket{b}\|}\right)^2$. This precision
is chosen to estimate the residual.
However, as seen in the numerical results of Sec.~\ref{sec:numerical_results},
the residuals are not small unless the algorithm stops, meeting 
the convergence criterion.
If high precision is unnecessary in the early stages of iterations,
the number of circuit runs can be reduced. 
Moreover, quantum amplitude estimation~\cite{brassard2002quantum,suzuki2020amplitude,grinko2021iterative}
can be applied within an acceptable circuit depth range.
These considerations suggest that optimizing the number of circuit runs and the circuit depth
while ensuring the algorithm's robustness improves the total query complexity.


In summary, our algorithm achieves a square root improvement for $\kappa$
concerning the maximum circuit depth, 
and the number of ancilla qubits is $n_a + 3$, regardless of both 
$\kappa$ and $\varepsilon$. Therefore, our algorithm fulfills
shallow circuits with constant ancilla qubits.
As a trade-off for the shallow circuits, the total query complexity is worse 
than that of other QLSAs. However, this can be improved by controlling 
the number of circuit runs and using the quantum amplitude estimation.

\section{Numerical results} \label{sec:numerical_results}
\begin{figure}[tbp]
	\centering
	\includegraphics[width=\linewidth]{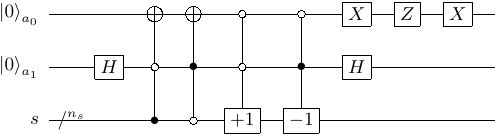}
	\caption{Circuit for $(1,2,0)$ block encoding of $A'=2\frac{A}{4}-I$.
					The $+1$ and $-1$ gates are defined in Fig.~\ref{fig:pm_1_gates}.}
	\label{fig:circuit_of_block_encoding_of_A}

	\centering
	\includegraphics[width=0.75\linewidth]{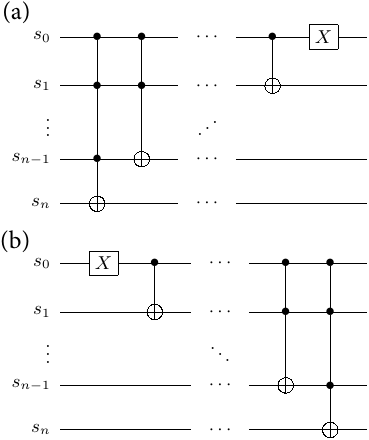}
	\caption{Gates that change the state by $\pm 1$. 
				  The circuit (a) represents the $+1$ gate for $n$ qubits. For $j=0,1,\dots,2^{n}-1$, this gate acts as 
					$\ket{j}_s\rightarrow \ket{j+1}_s$, where $\ket{2^n}_s\equiv\ket{0}_s$.
					The circuit (b) represents the $-1$ gate for $n$ qubits. For $j=0,1,\dots,2^{n}-1$, this gate acts as 
					$\ket{j}_s\rightarrow \ket{j-1}_s$, where $\ket{-1}_s\equiv\ket{2^n-1}_s$.}
	\label{fig:pm_1_gates}
\end{figure}

\begin{figure}[tbp]
  \includegraphics[width=\linewidth]{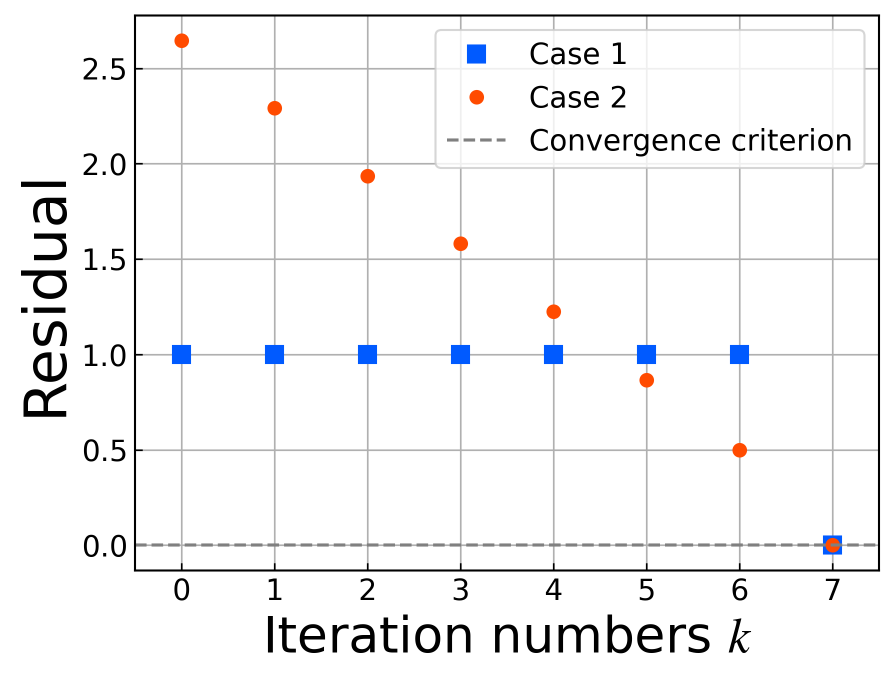}
  \caption{Plots of the residuals obtained in the simulations ($N=16$).
					 }
  \label{fig:residual_errors_without_shot_noises}
\end{figure}

We conduct numerical experiments on the Poisson equation to validate the QCG method 
using the QET. The one-dimensional Poisson equation is represented as
\begin{equation} \label{eq:poisson_equation}
  \frac{d^2 \phi}{d r^2} = -\rho(r),
\end{equation}
where $\rho$ is a given function, and $\phi$ is an unknown function.
We discretize the above equation using the second-order central difference scheme
with the Dirichlet boundary conditions $\rho(r_L)=\rho(r_R)=0$. This results in 
a linear system of equations $A\ket{x}=\ket{b}$ for
\begin{equation}\label{eq:A_of_poisson}
  A=\begin{bmatrix} 
    2      & -1     & 0      & \cdots  & 0 \\
    -1     & 2      & -1     & \cdots  & 0 \\
    0      & -1     & \ddots & \ddots  & \vdots\\
    \vdots & \vdots & \ddots & 2       & -1\\
    0      & \cdots & 0      & -1      & 2
  \end{bmatrix},
\end{equation}
\begin{align}
  \ket{x} &= \left[
    \phi_0, \phi_1, \cdots , \phi_{N-1}
  \right]^\top, \\
	\ket{b} &= \left[
    \rho_0 (\Delta x)^2, \rho_1 (\Delta x)^2, \cdots , \rho_{N-1} (\Delta x)^2
  \right]^\top,
\end{align}
where $N$ is the system size and $\Delta x = \frac{r_R-r_L}{N+1}$.
Since the matrix $A$ is a positive definite Hermitian matrix, 
the QCG method can solve this system.
Figure~\ref{fig:circuit_of_block_encoding_of_A} shows the circuit for 
$(1,2,0)$ block encoding of $A'=2\frac{A}{4}-I$ ($\alpha=4$). The $+1$ and $-1$ gates
in this circuit are presented in Fig.~\ref{fig:pm_1_gates}.
It is important to note that the subnormalization factor must satisfy $\alpha\geq 4$ 
because the maximum eigenvalue of $A$ is nearly $4$. Therefore,
the block encoding described above is nearly optimal in terms of the subnormalization factor.

We consider the system size $N=16$ and two cases for the initial vectors
\begin{align}
	\mathrm{Case}\ 1: &\quad \ket{b}=\frac{1}{\sqrt{2}}(\ket{N/2-1}+\ket{N/2}), \notag \\
	\mathrm{Case}\ 2: &\quad \ket{b}=\frac{1}{\sqrt{N}}\sum_{j=0}^{N-1}\ket{j}, \notag
\end{align}
which can be efficiently prepared using the Hadamard gates. 
Based on classical numerical calculations, the condition number is $\kappa\approx116$,
and the norm is $\|A\|\approx3.97$.
The error tolerance is $\varepsilon=0.1$, and the convergence criterion is 
$\||\ket{r_{k}}\|\leq \|A\|\varepsilon/\kappa\|\ket{b}\|=3.40\times10^{-3}$. 
States and their inner products are computed
using Statevector Simulator of Qiskt~\cite{Qiskit} as a backend. 
The QSPPACK~\cite{dong2021efficient,wang2022energy,dong2022infinite,Qsppack} is used
to calculate phase factors. 

\begin{table*}[t]
  \centering
  \caption{Comparison of polynomial degrees between the QCG method using the QET
					 and the QLSA using
					 the direct QSVT~\cite{martyn2021grand} for the one-dimensional Poisson equation.
					 Both cases have the same number of iterations.
					 Note that the condition number scales as $\kappa=O(N^2)$ for the system
					 size $N$.
					 }
    \begin{tabular}{ccccc} \hline \hline
      System size & Condition number & \multicolumn{3}{c}{Polynomial degree}  \\ 
			$N$ & $\kappa$ & QCG & Direct QSVT 	 & Rectangular function  \\ \hline 
			4   & 9.47 & 2  & $7.45\times10^{3}$ & $6.61\times10^{3}$ \\ 
			8   & 32.2 & 4  & $2.88\times10^{4}$ & $2.55\times10^{4}$ \\ 
			16  & 116  & 8  & $1.18\times10^{5}$ & $1.05\times10^{5}$ \\ 
			32  & 441  & 16 & $5.05\times10^{5}$ & $4.45\times10^{5}$ \\ \hline \hline
    \end{tabular}
  \label{tb:comparison_of_QCG_vs_direct_QSVT}
\end{table*}

\begin{figure*}[tbp]
  \centering
	\begin{minipage}[c]{0.45\linewidth}
		\centering
    \includegraphics[width=\linewidth]{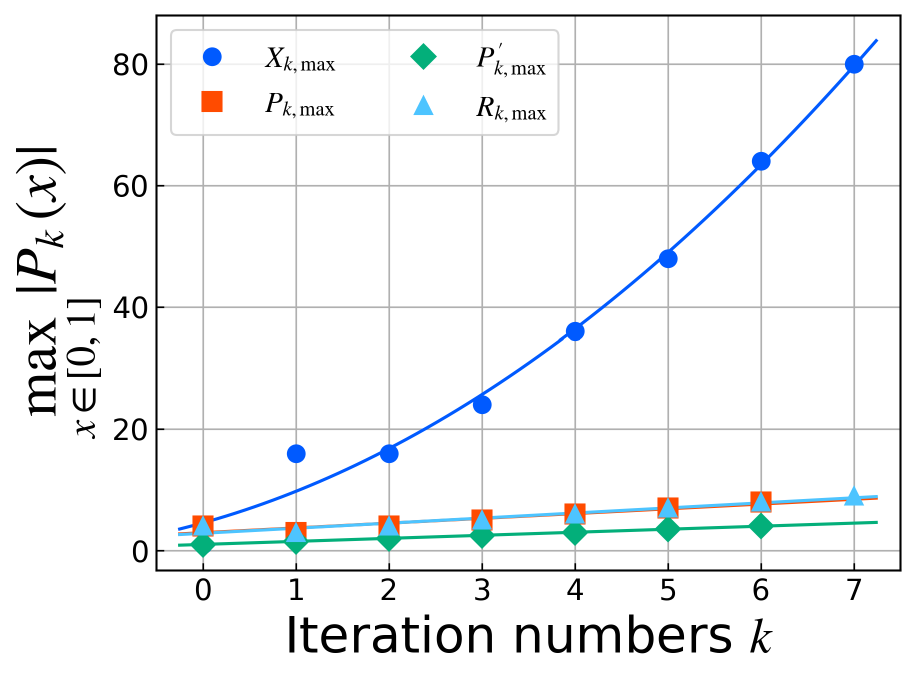}
		\put(-210,160){\Large{(a)}}
		\label{fig:max_of_polynomials_0_1_case1}
	\end{minipage} 
	\begin{minipage}[c]{0.45\linewidth}
		\centering
    \includegraphics[width=\linewidth]{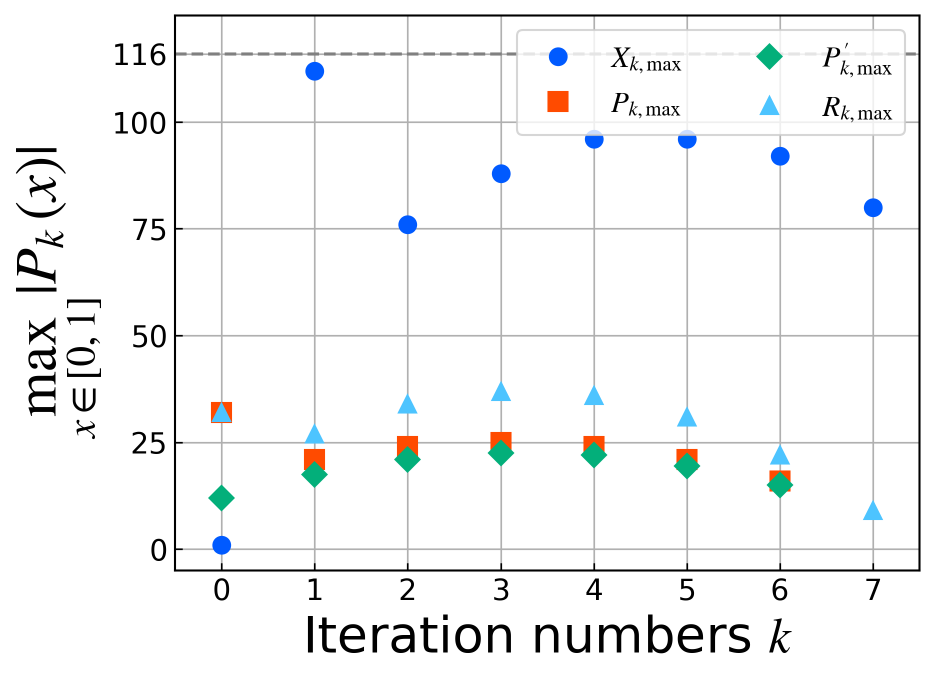}
		\put(-210,160){\Large{(b)}}
		\label{fig:max_of_polynomials_0_1_case2}
	\end{minipage}
  \caption{
		Plots of the maximum absolute values of polynomials for $x\in[0,1]$ ($N=16$): 
		(a) For Case 1 and (b) for Case 2. 
		The curve and straight lines in (a) represent
		the results of fitting quadratic and linear functions. The gray dashed line in (b) represents
		the condition number $\kappa\approx116$.
	}
  \label{fig:max_of_polynomials_0_1}
\end{figure*}

Figure~\ref{fig:residual_errors_without_shot_noises} shows the plots of the residuals.
As iterations progress, the residuals decrease, and the algorithm converges,
meeting the convergence criterion at the seventh iteration. 
At this point, the residual and error are shown below:
For Case 1,
\begin{align}
	\begin{split}
		\|\ket{r_8}\|&=1.89\times10^{-5}<3.40\times10^{-3}, \\ 
		\left\|\ket{x}-\frac{1}{\alpha}\ket{x_8}\right\|&=8.71\times10^{-5}<0.1, 
	\end{split} \notag
\end{align}
and for Case 2,
\begin{align}
	\begin{split}
		\|\ket{r_8}\|&=2.84\times10^{-5}<3.40\times10^{-3}, \\ 
		\left\|\ket{x}-\frac{1}{\alpha}\ket{x_8}\right\|&=2.71\times10^{-4}<0.1. 
	\end{split} \notag
\end{align}
These results demonstrate the successful resolution of the one-dimensional Poisson equation by our algorithm.

We compare the polynomial degrees between our algorithm and the QLSA using the direct QSVT~\cite{martyn2021grand} 
applied to the one-dimensional Poisson equation. Our algorithm determines the polynomial degree 
based on the number of iterations when the algorithm stops for $\varepsilon=10^{-1}$. 
For that number denoted by $m$, the maximum circuit depth is represented as $2(m+1)$.
On the other hand, in the QLSA using the direct QSVT,
the polynomial degree (equivalent to the maximum circuit depth)
is computed based on Eq.~\eqref{eq:degree_of_MI} for $\alpha=4$ and $\varepsilon=10^{-1}$.

Table~\ref{tb:comparison_of_QCG_vs_direct_QSVT} compares the degrees of these polynomials.
Notably, this result shows a substantial reduction in the polynomial degree (the maximum circuit depth) 
by our algorithm compared to the direct QSVT by three to four orders of magnitude.
The table also reveals that the polynomial degree of the direct QSVT 
is greatly influenced by the degree of the polynomial approximating the rectangular function, 
which consists of the sign function.
This finding indicates that approximating the sign function demands 
high polynomial degree.

Figure~\ref{fig:max_of_polynomials_0_1} plots the maximum absolute values 
of polynomials for $x\in[0,1]$ ($N=16$). The scalings of these values for Case 1 are
\begin{equation}
	X_{k, \max} = O(k^2), \quad P_{k, \max}, P'_{k, \max}, R_{k, \max} = O(k).
\end{equation}
On the other hand, the values for Case 2 depend on $\kappa$ rather than $k$.
Numerous circuit runs are required for both cases, as seen in the scaling of
Eq.~\eqref{eq:total_query_complexity_of_QCG_by_QET} or~\eqref{eq:total_query_complexity_of_QCG_by_QET2}
because these values are not close to 1.
Therefore, in the problem setting of the Poisson equation, the total query complexity
is enormous, although the circuit depth remains shallow.


\section{Discussion}
In the QLSAs, the parameters $N$ and $\kappa$ significantly impact achieving advantages
over the classical counterparts. 
Achieving these advantages requires $N$ to be significant and $\kappa$ to be small,
especially when $\kappa$ scales as $O(\mathrm{polylog}(N))$.
Our algorithm must also consider the dependence regarding
the maximum absolute values of the polynomials. Specifically, 
minimizing their dependence on $k$ or $\kappa$ is 
essential for reducing the number of circuit runs.
As shown in Fig.~\ref{fig:max_of_polynomials_0_1}, this dependence relies on 
the initial vector. Inherently, it can also depend on the given matrix $A$. 
Investigating the dependence concerning $A$ and $\ket{b}$ 
is a subject for future work.


The positive-side QET can be generalized when the Hermitian matrix is not positive semidefinite.
Suppose that we know in advance that the eigenvalues of $A/\alpha$ exist in $[\delta_1,\delta_2]$
for a Hermitian matrix $A$ and $-1\leq\delta_1<\delta_2\leq1$. We use the polynomial and matrix
\begin{align}
	P^{\mathrm{new}}(x) &= P\left(
		\frac{\delta_2-\delta_1}{2}x + \frac{\delta_1+\delta_2}{2}
	\right), \\
	A' &= \frac{2}{\delta_2-\delta_1} \frac{A}{\alpha} - \frac{\delta_1+\delta_2}{\delta_2-\delta_1}I.
\end{align}
Consequently, we eliminate the unnecessary region ($x\in[-1,\delta_1]\cup[\delta_2,1]$) of the 
original polynomial from the range $[-1,1]$. 
As discussed in Sec.~\ref{subsec:positive_side_QET},
we can efficiently construct the block encoding of $A'$ for a sparse matrix $A$.
Moreover, one can approximately build the block encoding of $A'$ from that of $A$ 
using the same procedure described in Appendix~\ref{appendix:explicit_circuits_for_block_encoding_of_of_A_dash}.

Although the circuit depth scaling of the QLSA using direct QSVT is nearly optimal,
the constant factor is significant. This is attributed to the high degree of the polynomial
used to approximate the sign function. This observation suggests that algorithms based on QSVT using approximate polynomials
of the sign function might face this issue. Additionally, calculating phase factors 
for polynomials with a high degree ($10^5\sim$) seems challenging~\cite{haah2019product,
chao2020finding,dong2021efficient,wang2022energy,ying2022stable,dong2022infinite}. 
Therefore, reducing the polynomial degree is necessary for both the circuit depth
and phase factor calculations.

The CG method works well for only positive definite Hermitian matrices. However, 
if we define $B=AA^{\dagger}$ for a general matrix $A$, then
$B$ becomes positive semidefinite and Hermitian. Accordingly, the solution takes the form
$\ket{x}=A^{\dagger}B^{-1}\ket{b}$, enabling the resolution of linear systems of equations
for general matrices through the QCG method using the QET. However,
the condition number of $B$ becomes $\kappa^2$, where $\kappa$ is the condition number of $A$.
As a result, the maximum circuit depth shows a linear dependence on $\kappa$.
This increase diminishes the advantage of the QCG method in terms of circuit depth.

The CG method is a form of the Krylov subspace method,
which is used for solving linear systems of equations 
and eigenvalue problems~\cite{shewchuk1994introduction,saad2003iterative}.
In particular, the Lanczos method is designed to create a Krylov subspace for Hermitian matrices. 
The vectors obtained through this method can also be 
represented as the product of a polynomial $P(A)$ and an initial vector $\ket{b}$
(as discussed in ~\cite{shao2018quantum}). Therefore, the Lanczos method can be implemented
using the QET. Similarly, considering $B=AA^{\dagger}$ for 
a general matrix $A$ allows the implementation of the Krylov subspace method in the QET framework.

\section{Summary}

This paper introduces the quantum conjugate gradient method using the
quantum eigenvalue transformation.
Additionally, we propose a new technique called the positive-side QET. 
This technique enables the implementation of polynomials with huge
absolute values on the negative side ($x\in[-1,0]$), which cannot be handled in the
previous QET framework.
The QCG method using the positive-side QET achieves a maximum circuit depth of 
$O\left(\sqrt{\kappa}\log\left(\frac{\kappa\|\ket{b}\|}{\|A\|\varepsilon}\right)\right)$.
This scaling represents a square root improvement
for $\kappa$ compared to earlier QLSAs~\cite{ambainis2010variable,
childs2017quantum,wossnig2018quantum,shao2018quantum,
gilyen2019quantum,lin2020optimal,martyn2021grand}. Moreover,
the number of ancilla qubits remains constant.
Therefore, our algorithm accomplishes shallow circuits with minimal ancilla qubits.
However, the algorithm demands numerous circuit runs, resulting in
an unfavorable total query complexity. 
To mitigate this cost, one can optimize a tradeoff between 
the precision for estimating inner products and the number of circuit runs,
a topic for future investigation.

\section*{Acknowledgements}
This work was supported by MEXT Quantum Leap Flagship Program Grants No. JPMXS0118067285 and No. JPMXS0120319794. K.W. was supported by JST SPRING, Grant No. JPMJSP2123.

\appendix
\renewcommand{\theequation}{A\arabic{equation} }
\setcounter{equation}{0}

\section{Explicit circuits for the block encoding of $A'$} \label{appendix:explicit_circuits_for_block_encoding_of_of_A_dash}
\begin{figure}[tbp]
	\centering
  \includegraphics[width=0.85\linewidth]{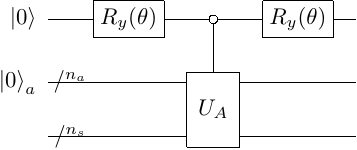}
  \caption{The circuit $U'_{A'}$ for the $(3,n_a+1,0)$ block encoding of $A'=2\frac{A}{\alpha}-I$.
					The angle is $\theta=2\tan^{-1}\sqrt{\frac{1}{2}}$.}
  \label{fig:circuit_of_block_encoding_of_A_dash}
\end{figure}

We discuss the explicit circuits constructing the block encoding of $A'=2A/\alpha-I$
from the block encoding of $A$. 
Let $A$ be a Hermitian matrix and $U_A$ be a $(\alpha,n_a,0)$ block encoding of $A$.
First, we construct the block encoding of the matrix $\frac{1}{\gamma} A'$ using
linear combination of unitaries, where $\gamma>1$ is an
amplification factor. Figure~\ref{fig:circuit_of_block_encoding_of_A_dash} shows
the circuit $U'_{A'}$ for the $(3,n_a+1,0)$ block encoding of $A'$.

Since $1/\gamma=1/3$ is unnecessary, this factor must be 
amplified to $1$. This amplification can be achieved by linear amplification~\cite{low2019hamiltonian,
gilyen2019quantum,martyn2021grand}, also called uniform spectral amplification.
Importantly, this method can be implemented using the QET (QSVT). The polynomial for the
linear amplification is represented as the product of $\gamma x$ and the approximate polynomial of 
the closed rectangular mentioned in Sec.~\ref{subsubsec:approximate_polynomials_of_sign_and_rectangular_functions}.
We present two methods for the linear amplification: One involving 
a gap $\Lambda$ related to the eigenvalues of $A$ and an alternative method without
the gap.

\subsection{Method with the gap $\Lambda$}
Let $\{\lambda_j\}$ be the eigenvalues of $A$ and $\Lambda\in(0,\frac{\alpha}{2})$ be 
a gap such that
\begin{equation}
  \Lambda \leq \lambda_{j} \leq \alpha-\Lambda.
\end{equation}
Therefore, we have
\begin{equation}
  -\frac{1-2\Lambda'}{\gamma}\leq \frac{2\frac{\lambda_{j}}{\alpha}-1}{\gamma}\leq\frac{1-2\Lambda'}{\gamma},
\end{equation}
where $\Lambda'=\frac{\Lambda}{\alpha}$.
For $\gamma>1$, one choice of the approximate polynomial for the closed rectangular function 
is represented as
$P_{\frac{1-\Lambda'}{\gamma},\frac{2\Lambda'}{\gamma},\frac{\varepsilon}{\gamma}}^{\mathrm{rect}}(x)$ such that
\begin{align}
  \begin{split}
    \left|1-P_{\frac{1-\Lambda'}{\gamma},\frac{2\Lambda'}{\gamma},\frac{\varepsilon}{\gamma}}^{\mathrm{rect}}(x)\right|&\leq\frac{\varepsilon}{\gamma},\\
        &x\in\left[-\frac{1-2\Lambda'}{\gamma},\frac{1-2\Lambda'}{\gamma}\right],\\
    \left|P_{\frac{1-\Lambda'}{\gamma},\frac{2\Lambda'}{\gamma},\frac{\varepsilon}{\gamma}}^{\mathrm{rect}}(x)\right|&\leq\frac{\varepsilon}{\gamma},\\
        &x\in\left[-1,-\frac{1}{\gamma}\right]\cup\left[\frac{1}{\gamma}, 1\right],\\
    \left|P_{\frac{1-\Lambda'}{\gamma},\frac{2\Lambda'}{\gamma},\frac{\varepsilon}{\gamma}}^{\mathrm{rect}}(x)\right|&\leq1,\quad x\in\left[-1,1\right].
  \end{split}
\end{align}
Then the polynomial for the linear amplification is defined by
\begin{equation}
	P_{\gamma,\alpha,\Lambda,\varepsilon}^{\mathrm{lamp}}(x)=\gamma xP_{\frac{1-\Lambda'}{\gamma},\frac{2\Lambda'}{\gamma},\frac{\varepsilon}{\gamma}}^{\mathrm{rect}}(x).
\end{equation}
This polynomial degree is expressed as
\begin{align}
  d_{\mathrm{lamp}}(\gamma,\alpha,\Lambda,\varepsilon)
    &= d_{\mathrm{rect}}\left(
      \frac{1-\Lambda'}{\gamma},\frac{2\Lambda'}{\gamma},\frac{\varepsilon}{\gamma}
    \right) + 1 \notag \\
    &= \mathcal{O}\left(
      \frac{\gamma\alpha}{\Lambda}\log\left(\frac{\gamma}{\varepsilon}\right)
    \right).
\end{align}

The polynomial $P_{\gamma,\alpha,\Lambda,\varepsilon}^{\mathrm{lamp}}(x)$ satisfies the following 
inequalities: For $|x|\leq\frac{1-2\Lambda'}{\gamma}$,
\begin{align}
  \left|\gamma x - P_{\gamma,\alpha,\Lambda,\varepsilon}^{\mathrm{lamp}}(x)\right| 
    &\quad\leq\gamma|x|\left|
      1-P_{\frac{1-\Lambda'}{\gamma},\frac{2\Lambda'}{\gamma},\frac{\varepsilon}{\gamma}}^{\mathrm{rect}}(x)
		\right| \notag \\
    &\quad\leq\gamma\cdot\frac{1-2\Lambda'}{\gamma}\cdot\frac{\varepsilon}{\gamma}\notag
    < \varepsilon.
\end{align}
For $|x|\leq\frac{1}{\gamma}$,
\begin{align}
  \left|P_{\gamma,\alpha,\Lambda',\varepsilon}^{\mathrm{lamp}}(x)\right| 
    &\leq\gamma|x|\left|P_{\frac{1-\Lambda'}{\gamma},\frac{2\Lambda'}{\gamma},\frac{\varepsilon}{\gamma}}^{\mathrm{rect}}(x)\right| \notag \\
    &\leq\gamma\cdot\frac{1}{\gamma}=1.
\end{align}
For $\frac{1}{\gamma}\leq|x|\leq1$,
\begin{align}
  \left|P_{\gamma,\alpha,\Lambda',\varepsilon}^{\mathrm{lamp}}(x)\right|
    &\leq\gamma|x|\left|P_{\frac{1-\Lambda'}{\gamma},\frac{2\Lambda'}{\gamma},\frac{\varepsilon}{\gamma}}^{\mathrm{rect}}(x)\right| \notag \\
    &\leq\gamma\cdot\frac{\varepsilon}{\gamma}=\varepsilon.
\end{align}

\begin{table*}[t]
  \centering
  \caption{Comparison of the methods for constructing the block encoding of $A'$.}
    \begin{tabular}{ccc} \hline \hline
      Method         				   & With $\Lambda$ & Without $\Lambda$ \\ \hline 
      \begin{tabular}{c}
				Approximate\\polynomial
			\end{tabular}            & $3 xP_{\frac{1-\Lambda/\alpha}{3},\frac{2\Lambda}{3\alpha},\frac{\varepsilon}{3}}^{\mathrm{rect}}(x)$ & $\frac{1}{1+\varepsilon/2}3 xP_{\frac{1+\varepsilon/4}{3},\frac{\varepsilon}{6},\frac{\varepsilon}{6}}^{\mathrm{rect}}(x)$ \\ 
      Query complexity 			   & $O\left(\frac{\alpha}{\Lambda}\log(1/\varepsilon)\right)$ & $O\left(\frac{1}{\varepsilon}\log(1/\varepsilon)\right)$\\         
      Requirement   				   & \begin{tabular}{c}
         The gap $\Lambda$ is large and \\the subnormalization factor $\alpha$ is small.
     \end{tabular} & The precision $\varepsilon$ is not small. \\ \hline \hline        
    \end{tabular}
  \label{tb:comparison_of_methods_for_constructing_block_encoding_of_A_dash}
\end{table*}

We denote the spectral decomposition of a Hermitian matrix $B$ by
\begin{equation}
  B = \sum_{j}\lambda_j\ket{\lambda_j}\bra{\lambda_j},
\end{equation}
where $\{\lambda_j\}$ is the eigenvalues such that $|\lambda_j|\leq\frac{1-2\Lambda'}{\gamma}$
and $\{\ket{\lambda_j}\}$ is the eigenvectors. We also denote 
\begin{equation}
  \tilde{\lambda}_j = P_{\gamma,\alpha,\Lambda,\varepsilon}^{\mathrm{lamp}}(\lambda_j),
\end{equation}
and then the inequality
\begin{equation}
  |\gamma\lambda_j - \tilde{\lambda}_j| \leq \varepsilon
\end{equation}
holds. Therefore, we approximately have
\begin{align}
  P_{\gamma,\alpha,\Lambda,\varepsilon}^{\mathrm{lamp}}(B) 
    &=\sum_{j}\tilde{\lambda}_j\ket{\lambda_j}\bra{\lambda_j} \notag \\
    &\approx\gamma\sum_{j}\lambda_j\ket{\lambda_j}\bra{\lambda_j}
    = \gamma B.
\end{align}

We get the $(1,n_a+2,\varepsilon)$ block encoding of $A'$ by
applying the above linear amplification for $\gamma=3$ to the block encoding $U'_{A'}$ and
using the QET circuit in Fig.~\ref{fig:QSVT_circuit_for_polynomial_with_definite_parity}(b). 
The polynomial degree used in this method is 
\begin{equation}
  d_{\mathrm{lamp}}(3,\alpha,\Lambda,\varepsilon)=O\left(\frac{\alpha}{\Lambda}\log(1/\varepsilon)\right).
\end{equation}
Therefore, the block encoding of $A'$ is built by using $O(\frac{\alpha}{\Lambda}\log(1/\varepsilon))$ additional queries 
to $U_A$.

\subsection{Method without the gap $\Lambda$}

In case the gap $\Lambda$ is not used,
for $\gamma>1,\varepsilon\in(0,1)$,
the approximate polynomial for the closed rectangular function is represented as
$P_{\frac{1+\varepsilon/4}{\gamma},\frac{\varepsilon}{2\gamma},\frac{\varepsilon}{2\gamma}}^{\mathrm{rect}}(x)$ and satisfies
the inequalities
\begin{align}
  \begin{split}
    \Bigl|1-&P_{\frac{1+\varepsilon/4}{\gamma},\frac{\varepsilon}{2\gamma},\frac{\varepsilon}{2\gamma}}^{\mathrm{rect}}(x)\Bigr|\leq\frac{\varepsilon}{2\gamma},\quad x\in\left[-\frac{1}{\gamma},\frac{1}{\gamma}\right],\\
    \Bigl|&P_{\frac{1+\varepsilon/4}{\gamma},\frac{\varepsilon}{2\gamma},\frac{\varepsilon}{2\gamma}}^{\mathrm{rect}}(x)\Bigr|\leq\frac{\varepsilon}{2\gamma},\\
			&\qquad x\in\left[-1,-\frac{1+\varepsilon/2}{\gamma}\right]\cup\left[\frac{1+\varepsilon/2}{\gamma}, 1\right],\\
    \Bigl|&P_{\frac{1+\varepsilon/4}{\gamma},\frac{\varepsilon}{2\gamma},\frac{\varepsilon}{2\gamma}}^{\mathrm{rect}}(x)\Bigr|\leq1,\quad x\in\left[-1,1\right].
  \end{split}
\end{align}
Then the polynomial for the linear amplification is defined by
\begin{equation}
  P_{\gamma,\varepsilon}^{\mathrm{lamp}}(x)=\frac{1}{1+\varepsilon/2}\gamma xP_{\frac{1+\varepsilon/4}{\gamma},\frac{\varepsilon}{2\gamma},\frac{\varepsilon}{2\gamma}}^{\mathrm{rect}}(x).
\end{equation}
This polynomial degree is expressed as
\begin{align}
  d_{\mathrm{lamp}}(\gamma,\varepsilon)
    &= d_{\mathrm{rect}}\left(
      \frac{1+\varepsilon/4}{\gamma},\frac{\varepsilon}{2\gamma},\frac{\varepsilon}{2\gamma}
    \right) + 1 \notag \\
    &= O\left(
      \frac{\gamma}{\varepsilon}\log\left(\frac{\gamma}{\varepsilon}\right)
    \right).
\end{align}

The polynomial $P_{\gamma,\varepsilon}^{\mathrm{lamp}}(x)$ satisfies the following 
inequalities: For $|x|\leq\frac{1}{\gamma}$,
\begin{align}
  &\left|\gamma x - P_{\gamma,\varepsilon}^{\mathrm{lamp}}(x)\right| \notag \\
    &\leq\gamma|x|\left|
      1-\frac{1}{1+\varepsilon/2}P_{\frac{1+\varepsilon/4}{\gamma},\frac{\varepsilon}{2\gamma},\frac{\varepsilon}{2\gamma}}^{\mathrm{rect}}(x)
		\right| \notag \\
    &\leq\gamma\cdot\frac{1}{\gamma}\left(
      \frac{1}{1+\varepsilon/2}\left|
        1 - P_{\frac{1+\varepsilon/4}{\gamma},\frac{\varepsilon}{2\gamma},\frac{\varepsilon}{2\gamma}}^{\mathrm{rect}}(x)
      \right| + \frac{\varepsilon/2}{1+\varepsilon/2}
    \right) \notag \\
    &\leq \frac{\varepsilon/2\gamma}{1+\varepsilon/2} + \frac{\varepsilon/2}{1+\varepsilon/2}
    <\varepsilon,
\end{align} 
and 
\begin{align}
  \left|P_{\gamma,\varepsilon}^{\mathrm{lamp}}(x)\right|
    &\leq\frac{1}{1+\varepsilon/2}\gamma|x|\left|
      P_{\frac{1+\varepsilon/4}{\gamma},\frac{\varepsilon}{2\gamma},\frac{\varepsilon}{2\gamma}}^{\mathrm{rect}}(x)
		\right|\notag \\
    &\leq\frac{1}{1+\varepsilon/2}\gamma\cdot\frac{1}{\gamma}<1.
\end{align}
For $\frac{1}{\gamma}\leq|x|\leq \frac{1+\varepsilon/2}{\gamma}$,
\begin{align}
  \left|P_{\gamma,\varepsilon}^{\mathrm{lamp}}(x)\right|
    &\leq\frac{1}{1+\varepsilon/2}\gamma |x|\left|
      P_{\frac{1+\varepsilon/4}{\gamma},\frac{\varepsilon}{2\gamma},\frac{\varepsilon}{2\gamma}}^{\mathrm{rect}}(x)
		\right| \notag \\
    &\leq\frac{1}{1+\varepsilon/2}\gamma\cdot\frac{1+\varepsilon/2}{\gamma}
    =1.
\end{align}
For $\frac{1+\varepsilon/2}{\gamma}\leq|x|\leq 1$,
\begin{align}
  \left|P_{\gamma,\varepsilon}^{\mathrm{lamp}}(x)\right|
    &\leq\frac{1}{1+\varepsilon/2}\gamma |x|\left|
			P_{\frac{1+\varepsilon/4}{\gamma},\frac{\varepsilon}{2\gamma},\frac{\varepsilon}{2\gamma}}^{\mathrm{rect}}(x)
		\right| \notag \\
    &\leq\frac{1}{1+\varepsilon/2}\gamma \cdot\frac{\varepsilon}{2\gamma}<\varepsilon.
\end{align}

Applying the above linear amplification for $\gamma=3$ to the block encoding $U'_{A'}$, we have 
the $(1,n_a+2,\varepsilon)$ block encoding of $A'$. The polynomial degree used in this method is 
\begin{equation}
  d_{\mathrm{lamp}}\left(
    3,\varepsilon
  \right)=O\left(\frac{1}{\varepsilon}\log(1/\varepsilon)\right).
\end{equation}
Therefore, the block encoding of $A'$ is built by using $O(\frac{1}{\varepsilon}\log(1/\varepsilon))$ additional queries 
to $U_A$.

Table~\ref{tb:comparison_of_methods_for_constructing_block_encoding_of_A_dash} summarizes the methods 
with and without the gap $\Lambda$. Since the query complexity of the method with $\Lambda$ depends 
linearly on $\frac{\alpha}{\Lambda}$, a larger gap and a smaller subnormalization factor are 
more desirable. On the other hand, the query complexity of the alternative method depends
linearly on $\frac{1}{\varepsilon}$ instead of $\frac{\alpha}{\Lambda}$. Therefore, 
a small or unknown gap and a large subnormalization factor are acceptable.
However, the precision $\varepsilon$ must not be small, indicating that
the high-precision construction is difficult.

\bibliographystyle{quantum}
\bibliography{reference}

\end{document}